\documentstyle[pra,multicol,aps]{revtex} 

\input{epsf}
\tighten  

\begin{document}
\preprint{preprint THEP 97/6}
\draft
\title{Stochastic wave function approach to the calculation of multitime 
correlation functions of open quantum systems}

\author{ Heinz--Peter Breuer, Bernd Kappler and Francesco Petruccione }
\address{
Albert-Ludwigs-Universit\"at, Fakult\"at f\"ur Physik, \\
Hermann-Herder Stra{\ss}e 3, D--79104 Freiburg im Breisgau,
Federal Republic of Germany}
\date{\today}
\maketitle


\begin{abstract}
Within the framework of probability distributions on projective
Hilbert space a scheme for the calculation of multitime correlation
functions is developed. The starting point is the Markovian stochastic
wave function description of an open quantum system coupled to an
environment consisting of an ensemble of harmonic oscillators in
arbitrary pure or mixed states.  It is shown that matrix elements of
reduced Heisenberg picture operators and general time-ordered
correlation functions can be expressed by time-symmetric expectation
values of extended operators in a doubled Hilbert space.  This
representation allows the construction of a stochastic process in the
doubled Hilbert space which enables the determination of arbitrary
matrix elements and correlation functions.  The numerical efficiency
of the resulting stochastic simulation algorithm is investigated and
compared with an alternative Monte Carlo wave function method proposed
first by Dalibard et al. [Phys. Rev. Lett. {\bf 68}, 580 (1992)].  By
means of a standard example the suggested algorithm is shown to be
more efficient numerically and to converge faster.  Finally, some
specific examples from quantum optics are presented in order to
illustrate the proposed method, such as the coupling of a system to a
vacuum, a squeezed vacuum within a finite solid angle, and a thermal
mixture of coherent states.
\end{abstract}
\begin{multicols}{2}

%
%

\section{Introduction}
\label{intro}

In recent years several stochastic wave function methods have been developed
and used to describe the dynamics of open quantum systems 
\cite{MolmerPRL68,Carmichael,GardinerPRA46,ZollerPRA46,Gisin:92,Gisin:93,BP:QS2,BP:QS4}. 
All these 
approaches are based on the following idea: instead of solving the 
quantum master equation to obtain the time evolution of the reduced density 
matrix, an ensemble of pure states is propagated using a stochastic time 
evolution.
This method provides two major advantages: the individual sample paths
of the different realizations can -- in some situations -- be interpreted as 
the time evolution of an individual continuously monitored quantum system 
\cite{WisemanMilburnPRA93,WisemanPRA93,BP:QS7,BP:QS8}
and the scaling of the numerical performance with the system size 
is better for this approach 
\cite{MolmerPRL68,ZollerPRA46,Gisin:92,MolmerOpt,BP:QS10}.

A particular interesting situation considered in quantum optics is the coupling
of a system (e.~g., an atom or a cavity) to the continuum of modes of the 
electromagnetic field. Since a lot of theoretical and 
experimental effort is used to prepare the environment in certain well 
defined states which are not restricted to thermal mixtures (see, e.~g., 
\cite{Parkins:93,Zoller:96,Wu:87}), it is interesting 
to investigate the time evolution of an open quantum system which is coupled 
to an environment in an arbitrary state. One way to obtain the stochastic 
time evolution of the reduced system takes as its starting point the quantum 
master equation for the reduced density matrix. A stochastic wave function
is constructed on a phenomenological basis in the following way: 
start with an ensemble of pure states representing the initial density
matrix $\rho(t_0)=\rho_0$ and propagate each member of the ensemble using 
a stochastic time evolution which guarantees 
that the covariance matrix of the stochastic process is the reduced 
density matrix. This procedure ensures that expectation values
of system observables are calculated correctly.

In applications of open quantum systems one is also interested in 
multitime correlation functions such as $\langle A(t+\tau)B(t)\rangle$,
where $A(t+\tau)$ and $B(t)$ are Heisenberg operators, and the angular
brackets denote the ensemble average.  In the density matrix 
approach these quantities are usually determined 
by making use of the quantum regression theorem \cite {Walls,GardinerQN} 
which is based on 
the identity
\begin{eqnarray}
  \label{q_reg_eq}
&&  \langle A(t+\tau)B(t)\rangle\nonumber\\\
&&  \hspace*{3em}=\mbox{Tr}_{\mbox{\scriptsize sys}}\bigg\{AV(t+\tau,t)
    \Big\{BV(t,t_0)\left\{\rho_0\right\}\Big\}\bigg\},
\end{eqnarray}
where $\mbox{Tr}_{\mbox{\scriptsize sys}}$ denotes the trace over the system's degrees of 
freedom and $V(t,t_0)$ is the time evolution super operator of the 
corresponding
quantum master equation, and $A$ and $B$ are Schr\"odinger operators. 
Eq.~(\ref{q_reg_eq}) allows the following 
interpretation: for the calculation of $\langle A(t+\tau)B(t)\rangle$ 
start with the 
density matrix $\rho(t_0)=\rho_0$ and propagate to $\rho(t)$ using 
the quantum master 
equation. Then propagate the ``density matrix'' $B\rho(t)$ up to the time
$t+\tau$ and calculate the quantum mechanical expectation value of $A$ with
respect to this density matrix. However, since $B\rho(t)$ is, in general, not
a positive matrix (and hence not a true density matrix), it can not be 
expressed as an ensemble of pure
states and hence a direct generalization of this computational scheme
to the stochastic wave function approach is not possible. Instead, some
alternative computational schemes are proposed in the literature such as 
expressing time-ordered multitime correlation functions as the sum
of symmetric time-ordered correlation functions \cite{MolmerPRL68} which can 
be simulated directly, or using the Heisenberg picture of the quantum state 
diffusion model \cite{Gisin:Heis}.

In this article we will present a different approach to the derivation
of the stochastic time evolution which is formulated within the framework
of probability distributions on projective Hilbert space \cite{BP:QS2,BP:QS4}.
The starting point for this approach is not the quantum Master equation
for the reduced density matrix, but a microscopic model 
for the dynamics of the total system. This model is used to calculate
the unitary time evolution of the total system in second order perturbation 
theory, which leads to a Liouville equation for the corresponding probability 
distribution.
The equation of motion of the reduced system's probability distribution 
is obtained by invoking a specific reduction formula, which relies
upon the above stated condition that expectations of system
operators are calculated accurately, and performing the Markov 
approximation. This scheme allows the derivation of the stochastic time 
evolution of an open quantum system coupled to an environment which is in an 
almost arbitrary state. To be more precise, the only restrictions we impose
on the state of the environment concern the relative time scales of
the system's and the environment's time evolution. Note, that this
restriction is not fundamental but necessary in order to end up with a
Markovian dynamics of the reduced system. 

To construct a stochastic process that allows also the determination of
multitime correlation functions we shall follow in this article the 
following strategy:
multitime correlation functions are expressed in terms of 
matrix elements of reduced Heisenberg 
picture operators which 
can be written  as expectation values in a {\it doubled} Hilbert
space. This then enables the formulation of  a stochastic process in the 
doubled Hilbert space which fulfills the condition, 
that matrix elements of reduced 
Heisenberg picture operators and hence multitime correlation functions are 
calculated correctly.

This article is organized as follows: In Sect.~\ref{prob_dist} we
will briefly review the statistical description of quantum systems 
in terms of probability distributions on the underlying Hilbert space.
Sec.~\ref{deri} describes the 
derivation of the equation of motion of the reduced system's probability
density functional. We find that the time evolution of this probability 
density is governed by a Liouville--Master equation. The 
resulting stochastic process is thus a piecewise deterministic Markov process.
In Sec.~\ref{heis} we define the matrix elements of the reduced Heisenberg 
picture operators. We show that these matrix elements can be determined by 
simulating a stochastic process in a doubled Hilbert space which is again a
piecewise deterministic Markov process. 
The process defined in the doubled Hilbert space is used to
calculate arbitrary measurable multitime correlation functions, which is
shown in Sec.~\ref{mult_corr}. In order to illustrate the general theory
presented in this article we explicitly discuss some examples of 
quantum optical systems in Sec.~\ref{examples}.

%
%
\section{Probability distributions on Hilbert space}
\label{prob_dist}

The basis for a description of open quantum systems in terms of a stochastic 
wave function is the definition of a probability measure on the underlying
Hilbert space ${\cal H}$. In this article we will make use of the probability 
density functional $P[\psi,t]$ which is defined such that 
$P[\psi,t]D\psi D\psi^*$ is the probability of 
finding the state of the system in the volume element $D\psi D\psi^*$
around $\psi$ at time $t$ \cite{BP:QS2,BP:QS4}. The volume element 
of the Hilbert space is defined as
\begin{equation}
  \label{vol_el_eq}
  D\psi D\psi^*\equiv\prod_x\frac{i}{2}d\psi(x)d\psi^*(x),
\end{equation}
where $x$ labels a complete set of quantum numbers. Thus, the normalization of
$P[\psi,t]$ reads
\begin{equation}
\label{P_norm_eq}
  \int D\psi D\psi^*P[\psi,t]=1,
\end{equation}
where the integral extends over the Hilbert space ${\cal H}$. In order to 
be consistent with the general principles of quantum mechanics we have
to impose two further conditions on $P[\psi,t]$: i) since two state
vectors which only differ by a phase factor describe the same physical state,
we require that $P[\psi,t]$ does not depend on the phase of the state vector 
and ii)
we are only interested in normalized state vectors, so that the support of 
$P[\psi,t]$ is the unit sphere in the Hilbert space ${\cal H}$. 
The quantum mechanical expectation value of an arbitrary linear 
operator $A$ with respect to $P[\psi,t]$ is defined as
\begin{equation}
  \label{exp_op_def_eq}
  \langle\!\langle A\rangle\!\rangle_{P[\cdot,t]}\equiv
  \int D\psi D\psi^*\langle\psi|A|\psi\rangle P[\psi,t].
\end{equation}
For the composition of two statistically independent subsystems, which are 
described by two probability distributions $P_1$ and $P_2$ on the Hilbert 
spaces ${\cal H}_1$ and ${\cal H}_2$, respectively,
we further define the tensor product probability distribution $P[\Psi,t]$
on ${\cal H}={\cal H}_1\otimes{\cal H}_2$  as \cite{BP:QS2,BP:QS4}
\begin{eqnarray}
  \label{prod_def_eq}
  P[\Psi,t]&=&\int D\psi D\psi^*\int D\varphi D\varphi^*\nonumber\\
  &\times&\delta[\Psi-\psi\otimes\varphi]P_1[\psi,t]P_2[\varphi,t],
\end{eqnarray}
where $\delta[\cdot]$ is the functional delta distribution, and write as a
short hand notation $P[\cdot,t]=P_1[\cdot,t]\otimes P_2[\cdot,t]$.
The connection to the standard density matrix description of statistical
ensembles is made through the identity
\begin{equation}
  \label{dens_mat_eq}
  \rho_t(x,x')\equiv\int D\psi D\psi^*\langle x|\psi
   \rangle \langle\psi|x'\rangle P[\psi,t].
\end{equation}
Hence, the density matrix is the covariance matrix of the stochastic process.

%
%
\section{Derivation of the equation of motion for 
the reduced system's probability distribution}
\label{deri}

In this section we derive the equation of motion
for the probability distribution of the states of an open system.
To this end, we first describe briefly the class of models we want to treat 
(Sec.~\ref{Model}). We then introduce in Sec.~\ref{WCA} the weak coupling 
assumption which enables
 us to derive an expression for the conditional transition
probability $T[\psi, t+\tau|\psi_0,t_0]$, where $\psi$, $\psi_0$ are pure
states of the open system (Sec.~\ref{cond_prob}).  
This result will be further simplified by making the Markov approximation
(Sec.~\ref{MA}), and finally be used to obtain the desired equation
of motion (Sec.~\ref{eq_mo}).

%
%

\subsection{Description of the underlying model}
\label{Model}

Consider an open quantum system 
in the Hilbert space ${\cal H}_1$ with Hamiltonian $H_1$ interacting with
an environment consisting of harmonic oscillators, e.~g., the electromagnetic 
field modes. The pure states of the environment are elements of the Hilbert 
space ${\cal H}_2$ and the Hamiltonian $H_2$ of the environment is given by
\begin{equation}
  \label{bath_ham}
  H_2=\sum_k \omega_k  b_k^\dagger b_k,
\end{equation}
where we set $\hbar =1$ for simplicity. 
Throughout this article,
$b_k$ denotes the annihilation
operator for the field mode $k=(\omega_k, \hat{\mbox{\boldmath $k$}}, \lambda_k)$
 with frequency $\omega_k$, unit wave vector
$\hat{\mbox{\boldmath $k$}}$, and polarization index $\lambda_k$.

The initial condition for the state of the environment at time $t_0$ is
given by a probability distribution 
\begin{equation}
\label{env_init_eq}
P_2[\varphi, t_0]=\sum_\alpha p_\alpha\int_0^{2\pi} \frac{d\chi}{2\pi} 
 \delta[\varphi-e^{i\chi}\varphi_\alpha],
\end{equation}
where the $p_\alpha$ are positive, normalized weights, i.\,e.,
\begin{equation}
  \label{p_alpha_eq}
  p_\alpha\ge 0,\quad \sum_\alpha p_\alpha=1,
\end{equation}
and the $\varphi_\alpha$ are arbitrary normalized  states in ${\cal H}_2$.
It is important to note that we do not suppose that the $\varphi_\alpha$
are orthogonal, i.\,e., the initial condition Eq.~(\ref{env_init_eq}) also
includes arbitrary mixtures of coherent states. We further assume that the 
system under consideration and the environment
are statistically independent at time $t_0$, i.\,e., that the probability 
distribution of the total system factorizes at time $t_0$ (cf. 
Eq.~(\ref{prod_def_eq})). 

The interaction of the system with its environment is modeled via the 
interaction Hamiltonian 
\begin{equation}
  \label{int_ham}
  H_I=\sum_{i=1,2} A_i\otimes B_i,
\end{equation}
where the operators $A_i$ and $B_i$ are defined as
\begin{eqnarray}
  A_1&=&A, \quad A_2=A^\dagger,\nonumber\\
  B_1& =&  -i\sum_k g_k b_k^\dagger, \quad B_2= B_1^\dagger.
\end{eqnarray}
The $g_k$ are the (real) coupling constants.
For optical systems $A$ and $A^\dagger$ could be, for example, the
positive and negative frequency part of the dipole operator.
It is useful to assume that the $A_i$ are eigenoperators of $H_1$, 
i.\,e., $[H_1, A_i]=\omega_iA_i$ with $\omega_1=-\omega_s$, and 
$\omega_2=\omega_s$, where $\omega_s$ is the system frequency. This 
is no restriction since this can always 
be achieved by choosing appropriate linear combinations of $A$ and $A^\dagger$.
This type of coupling describes for example the 
interaction of a two level system or a harmonic oscillator with a quantized 
electromagnetic field in the rotating wave
approximation, which is of particular interest in quantum optics. 
Note that the restriction to this kind of coupling has only technical reasons.
The generalization of the presented theory to a coupling involving more
than one system operator $A$, which was considered for example in 
Ref. \cite{BP:QS4}, is straightforward. The Hamiltonian of the 
total system is
\begin{equation}
  \label{H_tot_eq}
  H=H_1+H_2+H_I.
\end{equation}

In order to simplify the discussion in Sec. \ref{MA}, where we state the 
conditions necessary to perform the Markov approximation, we will 
introduce at this point a decomposition of the Hilbert space of the 
environment into the Hilbert space ${\cal H}_{\mbox{\scriptsize B}}$ which is the state space of
 the bath and the Hilbert space ${\cal H}_{\mbox{\scriptsize Dr}}$ which describes the
 driving field. This decomposition reflects the fact that the 
electromagnetic field which interacts with the system is in general 
produced by different sources which can have totally different effects
on the system. This decomposition is defined as follows:

Let ${\cal K}_{\mbox{\scriptsize B}}$ be the set of modes
$k$ for which $\langle\!\langle b_k\rangle\!\rangle_{P_2[\cdot,t_0]}= 0$, i.\,e.,
the modes which do not contribute to the average electromagnetic field and 
 ${\cal K}_{\mbox{\scriptsize Dr}}$ be the set of modes
$k$ for which $\langle\!\langle b_k\rangle\!\rangle_{P_2[\cdot,t_0]}\neq 0$.
Then we define the Hilbert space of the bath ${\cal H}_{\mbox{\scriptsize B}}$ as
\begin{equation}
  \label{bath_Hil_eq}
  {\cal H}_{\mbox{\scriptsize B}}=\bigotimes_{k\in{\cal K}_{\mbox{\scriptsize B}}}{\cal H}_k,
\end{equation}
where ${\cal H}_k$ is the Hilbert space of the field mode $k$, and similarly 
we define the Hilbert space ${\cal H}_{\mbox{\scriptsize Dr}}$ of the driving field. 
Obviously we have ${\cal H}_2={\cal H}_{\mbox{\scriptsize B}}\otimes{\cal H}_{\mbox{\scriptsize Dr}}$. 
We further assume that the states belonging to ${\cal H}_{\mbox{\scriptsize B}}$ 
and ${\cal H}_{\mbox{\scriptsize Dr}}$ are statistically independent, i.\,e., the probability 
distribution of the environment can be written as $P_2=P_{\mbox{\scriptsize B}}\otimes
P_{\mbox{\scriptsize Dr}}$, where $P_{\mbox{\scriptsize B}}$ and $P_{\mbox{\scriptsize Dr}}$ are the probability distributions on the
bath space and on the space of the driving field respectively. This assumption 
is reasonable since the two fields are usually 
generated by different and independent sources.

With the above definitions in mind we can  decompose accordingly the 
operators $B_i$ as follows:
\begin{equation}
  \label{B_deco_eq}
  B_i=B_i^{\mbox{\scriptsize B}}\otimes I + I \otimes B_i^{\mbox{\scriptsize Dr}},
\end{equation}
where 
\begin{equation}
  \label{B_op_def_eq}
  B_1^{\mbox{\scriptsize B}}=-i\sum_{k\in{\cal K}_{\mbox{\scriptsize B}}}g_k b_k^\dagger, \quad B_2^{\mbox{\scriptsize B}}={B_1^{\mbox{\scriptsize B}}}^\dagger,
\end{equation}
and
\begin{equation}
  \label{B_Dr_op_def_eq}
  B_1^{\mbox{\scriptsize Dr}}=-i\sum_{k\in{\cal K}_{\mbox{\scriptsize Dr}}}g_k b_k^\dagger, \quad B_2^{\mbox{\scriptsize Dr}}=
  {B_1^{\mbox{\scriptsize Dr}}}^\dagger.
\end{equation}
For the expectations
of the operators $B_i$ we thus obtain
\begin{equation}
  \label{exp_eq}
  \langle\!\langle B_i\rangle\!\rangle_{P_2[\cdot,t]}=\langle\!\langle B_i^{\mbox{\scriptsize Dr}}\rangle\!\rangle_{P_{\mbox{\tiny Dr}}[\cdot,t]},
\end{equation}
i.\,e., the expectation of the electromagnetic field 
only depends on the state of the driving field and 
is independent of the state of the bath. For the two time correlation function
$ \langle\!\langle B_i^\dagger (s) B_j\rangle\!\rangle_{P_{2}[\cdot,t]}$,
where $B_i(s)=\exp(iH_2s)B_i\exp(-iH_2s)$, we find
\begin{eqnarray}
  \label{exp1_eq}
  \langle\!\langle B_i^\dagger(s)B_j\rangle\!\rangle_{P_2[\cdot,t]}&=&
  \langle\!\langle {B_i^{\mbox{\scriptsize B}}}^\dagger (s) B_j^{\mbox{\scriptsize B}}
  \rangle\!\rangle_{P_{\mbox{\tiny B}}[\cdot,t]}\nonumber\\
  &+&  \langle\!\langle {B_i^{\mbox{\scriptsize Dr}}}^\dagger (s)B_j^{\mbox{\scriptsize Dr}}
  \rangle\!\rangle_{P_{\mbox{\tiny Dr}}[\cdot,t]},
\end{eqnarray}
which means that this correlation functions can be split into the sum of 
two terms, 
where the first term only depends on the state of the bath and the second
term is due to the driving field.

%
%

\subsection{Dynamics of the environment -- weak coupling assumption}
\label{WCA}

The basis of the weak coupling assumption is the fact
that the environment is large compared
to the system. Hence it is assumed that the dynamics of the environment is 
given 
by its free time evolution; the perturbations due to the interaction with the  
system under 
consideration can be neglected. For the probability distribution
of the environment this means
\begin{equation}
  \label{bath_mo}
  P_2\left[\varphi,t\right]=\sum_\alpha p_\alpha\int_0^{2\pi}\frac{d\chi}{2\pi}
  \delta\left[\varphi-e^{i\chi}e^{-iH_2(t-t_0)}\varphi_\alpha\right]
\end{equation}
for all $t$,
and we have chosen the initial condition given in Eq.~(\ref{env_init_eq}).

Using the weak coupling assumption, we can further evaluate
the expressions for the bath correlation functions
$\langle\!\langle {B_i^{\mbox{\scriptsize B}}}^\dagger (s) B_j^{\mbox{\scriptsize B}}\rangle\!\rangle_{P_{\mbox{\tiny B}}[\cdot,t]}$ 
defined in Sec.~\ref{Model}, if the states of the {\it individual bath modes}
 are statistically independent, 
which means that the probability distribution of the bath is the product
of probability distributions $P_k$ on the Hilbert spaces ${\cal H}_k$ of the
individual bath modes $k$. This will be the case for the examples we want to 
treat in this article. We obtain for example
\begin{equation}
  \label{bath_correl_eq}
  \langle\!\langle {B_2^{\mbox{\scriptsize B}}}^\dagger (s) B_2^{\mbox{\scriptsize B}}\rangle\!\rangle_{P_{\mbox{\tiny B}}[\cdot,t]}
  =\sum_{k\in{\cal K}_{\mbox{\scriptsize B}}} g_k^2e^{i\omega s}
  \langle\!\langle  b_k^\dagger b_k\rangle\!\rangle_{P_{k}[\cdot,t_0]}.
\end{equation}
It is important to note, that this correlation function only depends on the 
time argument $s$ and the state of the bath at time $t_0$; it is independent of
 the time $t$.

%
%

\subsection{Dynamics of the reduced system}

In this section we focus on the description of the dynamics of 
the reduced system evolving from the initial distribution of the total system
\begin{equation}
  \label{P_in_eq}
  P[\cdot,t_0]=P_1[\cdot,t_0]\otimes  P_2[\cdot,t_0]
\end{equation}
(cf. Eq.~(\ref{prod_def_eq}) for the definition of the tensor product 
probability distribution), where
\begin{equation}
  \label{P_1_in_eq}
  P_1[\psi,t_0]=\int_0^{2\pi} \frac{d\chi}{2\pi} \delta[\psi-e^{i\chi}\psi_0],
\end{equation}
and $P_2[\varphi,t_0]$ is given by Eq.~(\ref{env_init_eq}).
Thus at time $t_0$ the system is in the pure 
state $\psi_0$ and the initial distribution of the environment is given by 
the probability distribution $P_2$. The dynamics of the reduced system is 
completely described through the 
conditional transition probability $T[\psi,t_0+\tau|\psi_0,t_0]$, 
which is the probability that the system is in 
the state $\psi$ at time $t_0+\tau$ {\it under the condition} that the
system is in the state $\psi_0$ at time $t_0$. In order to do this, we 
will start in Sec.~\ref{pure_sep_dyn}  with the 
calculation of the time evolution of the pure product state  
$\Psi_\alpha(t_0)=\psi_0\otimes \varphi_\alpha$ in second order
perturbation theory. Using this result we can calculate $P[\Psi,t_0+\tau]$
 and applying a reduction formula
(Sec.~\ref{cond_prob}), we finally obtain an exact expression for 
the conditional transition probability.  

It is important to note that the conditional transition probability 
$T[\psi,t_0+\tau|\psi_0,t_0]$ also defines the dynamics of the reduced system, 
if its initial state is given by an arbitrary probability distribution
$P_1[\psi,t_0]$. According to the general laws of probability theory
the propagated reduced probability distribution $P_1[\psi,t_0+\tau]$ given
the conditional transition probability $T$ is
\begin{equation}
\label{P_prop_1_eq} 
   P_1[\psi,t_0+\tau]=\int D\psi_0 D\psi_0^*T[\psi,t_0+\tau|\psi_0,t_0]
   P_1[\psi_0,t_0].
\end{equation}
Thus it is sufficient to consider the initial conditions Eq.~(\ref{P_1_in_eq}).

%
%

\subsubsection{Time evolution of the total system in second order
  perturbation theory}

\label{pure_sep_dyn}
In this section we will study the time evolution of the initial distribution
Eq.~(\ref{P_in_eq}) in second order perturbation theory. To this end, we 
have to calculate the time evolution of the pure product state  
$\Psi_\alpha(t_0)=\psi_0\otimes \varphi_\alpha$.
In the interaction picture which coincides with the Schr\"odinger picture
at time $t_0$, the operators $A_i(\tau)$ are given by 
\begin{equation}
  \label{A_eq}
  A_i(\tau)=A_ie^{i\omega_i\tau}, \quad i=1,2,
\end{equation}
since $A_i$ is an eigenoperator of $H_1$, and the operators
$B_i(\tau)$ are given by 
\begin{eqnarray}
  \label{B_eq}
  B_1(\tau)&=&-i\sum_kg_k b_k^\dagger e^{i\omega_k\tau},\nonumber\\
  B_2(\tau)&=&i\sum_kg_k b_ke^{-i\omega_k\tau}.
\end{eqnarray}
In second order perturbation theory we obtain for the propagated state
\begin{eqnarray}
  |{\Psi_\alpha(t_0+\tau)}\rangle &=&|{\psi_0}\rangle\otimes 
  |{\varphi_\alpha}\rangle+\sum_{i} A_i|{\psi_0}\rangle
  \otimes f_i|{\varphi_\alpha}\rangle\nonumber\\
  \label{pert}
  &+&\sum_{i,j} A_i^\dagger A_j|{\psi_0}\rangle\otimes g_{ij}
  |{\varphi_\alpha}\rangle,
\end{eqnarray}
where the first and second order propagation operators acting on the states
of the environment are
\begin{eqnarray}
\label{f_eq}  
  f_i&=&-i\int_0^\tau ds\; e^{i\omega_{i}s} B_i(s),\\
\label{g_eq} 
  g_{ij}&=&-\int_0^\tau ds\int_0^{\tau-s}ds'
e^{-i\omega_{i}(s+s')}e^{i\omega_{j}s}\nonumber\\
&& \times B_i^\dagger(s+s'){B_j}(s).
\end{eqnarray}
If we define the averages of $f_i$ and $g_{ij}$ with respect to the 
probability distribution of the environment as
\begin{equation}
\label{F_G_eq}
F_i\equiv\langle\!\langle f_i\rangle\!\rangle_{P_2[\cdot,t_0]},\quad
G_{ij}\equiv\langle\!\langle g_{ij}\rangle\!\rangle_{P_2[\cdot,t_0]},
\end{equation}
and the shifted operators $\widetilde f_i$ and $\widetilde g_{ij}$ as
\begin{eqnarray}
  \label{f_g_eq}
  \widetilde{f}_i\equiv  f_i-F_i,\quad 
 \widetilde{g}_{ij}\equiv  g_{ij}-G_{ij},
\end{eqnarray}
we can rewrite Eq.~(\ref{pert}) as
\begin{eqnarray}
  \label{Pert2_eq}
  |{\Psi_\alpha(t_0+\tau)} \rangle&=&L|\psi_0\rangle\otimes |{\varphi_\alpha}\rangle
  +\sum_{i} A_i|{\psi_0}\rangle\otimes \widetilde{f}_i
  |{\varphi_\alpha}\rangle\nonumber\\
  &+&\sum_{i,j} A_i^\dagger A_j|{\psi_0}\rangle\otimes\widetilde{g}_{ij}
  |{\varphi_\alpha}\rangle
\end{eqnarray}
where we introduced the time evolution operator $L$ which is defined as 
\begin{equation}
  \label{L_eq}
  L=I+\sum_i F_i A_i+\sum_{i,j}G_{ij} A_i^\dagger A_j.
\end{equation}
As we will show in Sec.~\ref{cond_prob}, Eq.~(\ref{Pert2_eq}) is the basis for
the construction of a specific reduction formula. 

We can also use Eq.~(\ref{Pert2_eq}) to determine the time evolution of the 
probability distribution of the total 
system if the initial probability distribution is given by 
Eq.~(\ref{P_in_eq}).
For the propagated distribution $P[\Psi,t_0+\tau]$ we simply obtain  
\begin{equation}
  \label{P_prop_eq}
   P[\Psi,t_0+\tau]=\sum_\alpha p_\alpha\int_0^{2\pi}\frac{d\chi}{2\pi}
   \delta[\Psi-e^{i\chi}\Psi_\alpha(t_0+\tau)], 
\end{equation}
where the states $\Psi_\alpha(t_0+\tau)$ are the propagated
states defined in Eq.~(\ref{Pert2_eq}).

%
%

\subsubsection{Reduction formula}
\label{cond_prob}

The key to the stochastic description of open quantum systems is the
reduction formula \cite{BP:QS2,BP:QS4}. The reduction formula states 
how the 
conditional transition probability $T[\psi,t_0+\tau|\psi_0,t_0]$
is defined given the probability distribution $P[\Psi,t]$ 
(cf. Eq.~(\ref{P_prop_eq})) on 
the state space of the total system ${\cal H}_1\otimes{\cal H}_2$.
In order to derive an appropriate expression for $T$ we require the following
necessary condition to hold:
\begin{eqnarray}
  \label{exp_cond_full_eq}
  &&\int D\Psi D\Psi^* \langle\Psi|C\otimes I|\Psi\rangle 
  P[\Psi,t_0+\tau]\nonumber\\
  &\stackrel{!}{=}&\int D\psi D\psi^* \langle\psi|C|\psi\rangle 
  T[\psi,t_0+\tau|\psi_0,t_0],
\end{eqnarray}
or in the more compact notation introduced in Sec.~\ref{prob_dist}
\begin{equation}
  \label{exp_cond_eq}
  \langle\!\langle C\otimes I\rangle\!\rangle_{P[\cdot,t_0+\tau]}
  \stackrel{!}{=}\langle\!\langle C 
  \rangle\!\rangle_{T[\cdot,t_0+\tau|\psi_0,t_0]}.
\end{equation}
Note that the integral on the left-hand side of Eq.~(\ref{exp_cond_full_eq})
extends over the Hilbert space ${\cal H}_1\otimes {\cal H}_2$, whereas
the integral on the right-hand side only extends over the reduced state space
${\cal H}_1$.
The above condition ensures that the time evolution of the expectation value 
of an arbitrary operator $C\otimes I$ acting on the total system's state 
space  
calculated using the distribution $P[\Psi,t_0+\tau]$ is the same as
the expectation value of $C$ with respect to $T[\psi,t_0+\tau|\psi_0,t_0]$.
However, there are infinitely many ways to construct a reduced 
probability distribution which fulfills the above condition, 
just as there are infinitely many ways to express an
arbitrary  density matrix as an ensemble of pure states \cite{HJW:rhoensemble},
and each reduced probability distribution will define a different stochastic 
process. 
In Ref. \cite{BP:QS8} it was shown that different processes can be
formulated using different ``reduction bases'' (which are bases of the
Hilbert space of the environment). These processes were interpreted as 
the time evolution of a continuously monitored individual quantum system
and each process was related to a specific measurement scheme 
\cite{WisemanPRA93}.

Here, we will follow a different approach, which has the advantage that 
it results in a particular simple, basis-free reduction formula, 
but which is not based on a specific measurement scheme.
If we insert the propagated probability distribution of the total system 
Eq.~(\ref{P_prop_eq}) in the left-hand side of the consistency condition
Eq.~(\ref{exp_cond_eq}) and neglect terms of the order $g_k^3$ we obtain 
\begin{eqnarray}
  \label{exp_prop}
  \langle\!\langle C\otimes I\rangle\!\rangle_{P[\cdot,t_0+\tau]}&=&\langle{\psi_0}|L^\dagger C L|{\psi_0}\rangle\nonumber\\
&&+ \sum_{i,j}\Gamma_{ij}\langle{\psi_0}| A_i^\dagger C A_j|{\psi_0}\rangle,
\end{eqnarray}
where we introduced the $(2\times 2)$--correlation matrix of the environment 
\begin{equation}
  \label{Gamm_eq}
  \Gamma_{ij}= \langle\!\langle \widetilde{f}_i^\dagger\widetilde{f}_j\rangle\!\rangle_{P_2[\cdot,t_0]}
  =-(G_{ij}+G_{ji}^*+F_i^*F_j),
\end{equation}
and $F_i$ and $G_{ij}$ are defined in Eq.~(\ref{F_G_eq}).
Note that $\Gamma$ is Hermitian and non-negative. If we denote the 
normalized eigenvectors of $\Gamma$ by ${\mu}_i
=(\mu_{1i},\mu_{2i})^T$, and its eigenvalues by $\widetilde\lambda_i$, and
define the system (jump-) operators as
\begin{equation}
\label{J_op_eq}
J_i=\sum_k\mu_{ki}^*A_k, 
\end{equation}
then Eq.~(\ref{exp_prop}) can be written as
\begin{eqnarray}
  \label{exp_prop_dia}
  \langle\!\langle C\otimes I\rangle\!\rangle_{P[\cdot,t_0+\tau]}&=&
  \langle{\psi_0}|L^\dagger C L|{\psi_0}\rangle\nonumber\\
  &&+ \sum_{i}\widetilde{\lambda}_i\langle{\psi_0}| J_i^\dagger C 
  J_i|{\psi_0}\rangle\nonumber\\
  &\stackrel{!}{=}&\langle\!\langle C 
  \rangle\!\rangle_{T[\cdot,t_0+\tau|\psi_0,t_0]}.
\end{eqnarray}
Consider the following conditional transition probability:
\begin{eqnarray}
  \label{T_eq}
  T[\psi,t_0+\tau|\psi_0,t_0]&=&w\delta[\psi- w^{-1/2}L\psi_0]\nonumber\\
  &+&\sum_{i}\tilde{\lambda}_i
  w_i\delta[\psi-w_i^{-1/2}J_i\psi_0],
\end{eqnarray}
where $w=||L{\psi_0}||^2$, and $w_i=||J_i{\psi_0}||^2$.
Obviously, the conditional transition probability defined by Eq.~(\ref{T_eq})
satisfies the necessary condition Eq.~(\ref{exp_cond_eq}). 
Since in the generic case the matrix $\Gamma$ is nondegenerated we find 
three contributions to $T$: the system can evolve into the state 
$w^{-1/2}L\psi_0$, or to one of the two states $w_1^{-1/2}J_1\psi_0$, and 
$w_2^{-1/2}J_2\psi_0$.
Note, that the expression for the conditional transition probability
Eq.~(\ref{T_eq}) is {\it exact} within second order perturbation theory;
however, its derivation is based on the special choice of the initial 
probability distribution Eq.~(\ref{P_in_eq}) and the complete dynamics
of the environment is contained in the functions $w$, $w_i$, 
$\widetilde\lambda_i$, and the operators $L$, and $J_i$.

%
%

\subsection{The Markov approximation}
\label{MA}

In Sec.~\ref{cond_prob} we derived an expression for the conditional
transition probability $T[\psi,t_0+\tau|\psi_0,t_0]$ (cf. Eq.~(\ref{T_eq}))
which is exact within second order perturbation theory.
However, in situations where the environment has a large number of degrees
of freedom, this formula is of little practical use for computations 
without further
approximations. The simplest approximation scheme resulting in a
conditional transition probability which is independent of the actual
state of the environment, is the Markov approximation. We will use
this approximation to calculate the quantities involving the state of the
environment, namely the ensemble average of the first order propagation
operator $F_i$ (cf.~Eqs.~(\ref{f_eq}) and~(\ref{F_G_eq})) and the ensemble 
average of the second order propagation operator $G_{ij}$ 
(cf.~Eqs.~(\ref{g_eq}), and~(\ref{F_G_eq})). This will be done in 
Sec.~\ref{F_calc} and Sec.~\ref{G_calc}, respectively.  
These results will then be combined in Sec.~\ref{T_MA_sec} in order
to establish an approximate expression for the conditional transition 
probability.

%
%

\subsubsection{Calculation of the ensemble average of the first order 
propagation operator}

\label{F_calc}
We can simply calculate the ensemble average $F_i$ of the first order 
propagation operator $f_i$ (cf. Eq.~(\ref{f_eq})) by Taylor expansion. 
To first order in $\tau$ we obtain:
\begin{equation}
  \label{F_app_eq}
  F_1=-\tau {\cal E}_{t_0}^*, \quad 
  F_2= \tau {\cal E}_{t_0},
\end{equation}
where ${\cal E}_{t_0}$ is the mean electromagnetic field at time $t_0$, 
weighted 
with respect to the coupling constants $g_k$, and the probability distribution 
$P_2[\cdot,t_0]$. Using Eq.~(\ref{exp_eq}) we find
\begin{equation}
  \label{E_t_eq}
  {\cal E}_{t_0}=-i\langle\!\langle B_2\rangle\!\rangle_{P_{2}[\cdot,t_0]}=
  \sum_{k\in{\cal K}_{\mbox{\scriptsize Dr}}}g_k
  \langle\!\langle b_k\rangle\!\rangle_{P_{\mbox{\tiny Dr}}[\cdot,t_0]}.
\end{equation}
Obviously, the quantities $F_i$ only depend on the driving field and are 
independent of the state of the bath. However, the validity
of this approximation is limited to the validity of the Taylor expansion 
to first order in $\tau$, and this is guaranteed by the condition
\begin{equation}
  \label{tau_tau_E_cond}
  \tau\ll\tau_{\cal E},
\end{equation}
where $\tau_{\cal E}$ is the correlation time of the mean electromagnetic
field; this correlation time is of the order of magnitude of the inverse 
bandwidth of the spectrum of the driving field.

%
%

\subsubsection{Calculation of the ensemble average of the second order
propagation operator}

\label{G_calc}
Performing a Taylor expansion of Eq.~(\ref{g_eq}) with respect to $\tau$ in
order to calculate the ensemble average of the second order
propagation operator $G_{ij}$ we are lead to a
vanishing first order term, i.\,e., $G_{ij}$ is at least quadratic in 
$\tau$. However, this Taylor expansion is only valid for a propagation time 
$\tau$ which is small compared to the decay time of the correlation function
of the environment $\langle\!\langle B_i^\dagger(s)B_j
\rangle\!\rangle_{P_2[\cdot,t_0]}$.
Since this correlation function is the sum of the correlation function of 
the bath and the driving field (cf. Eq.~(\ref{exp1_eq})), we find that 
$G_{ij}$ is the sum of two terms, where the first one depends only on the 
state of the bath, and the second depends only on the state of the driving 
field, i.\,e., 
\begin{equation}
\label{G_sum_eq}
G_{ij} = G_{ij}^{\mbox{\scriptsize B}} +G_{ij}^{\mbox{\scriptsize Dr}}.
\end{equation}
Since we focus our interest on situations, where the decay time $\tau_{\mbox{\scriptsize B}}$ 
of the bath correlation 
is very small compared to the decay time $\tau_{\mbox{\scriptsize Dr}}$ of the driving field 
correlation (which is equal to $\tau_{\cal E}$ for a coherent driving field) 
and to the time scale $\tau_s$ of
the systematic dynamics, we have to use an appropriate approximation 
of $G_{ij}$ for ``medium'' values of $\tau$, i.\,e., for values of $\tau$ for 
which
\begin{equation}
\label{Mark_cond_eq}
\tau_{\mbox{\scriptsize B}}\ll\tau\ll\tau_{\mbox{\scriptsize Dr}},\tau_{\cal E}, \tau_s.
\end{equation}
In this case, the bath correlation function leads to a contribution 
$G_{ij}^{\mbox{\scriptsize B}}$ linear in $\tau$ whereas the driving field correlation 
function leads to a term $G_{ij}^{\mbox{\scriptsize Dr}}$ which is quadratic in $\tau$; 
thus we will neglect the latter. By describing the dynamics
of the open system on a {\it coarse grained} time axis, i.\,e., by neglecting
time variations which are smaller than $\tau_{\mbox{\scriptsize B}}$ we can perform the 
Markov approximation
and extend the range of the integration over $s'$ in Eq.~(\ref{g_eq}) to 
infinity. The result to first order in $\tau$ is
\begin{equation}
  \label{G_lin_eq}
  G_{ij}=-\tau\int_0^\infty ds\; e^{-i\omega_i s} 
  \langle\!\langle {B_i^{\mbox{\scriptsize B}}}^\dagger (s) B_j^{\mbox{\scriptsize B}}\rangle\!\rangle_{P_{\mbox{\scriptsize B}}[\cdot,t_0]}.
\end{equation}
We can simplify this further by evaluating the above integral, as shown
in Appendix ~\ref{G_ij_app}. The result is:
\begin{eqnarray}
  \label{G_11_0_eq}
  G_{11}&=&-\tau\left(\frac{\gamma}{2}\left(N+1\right)
  -i\left(S_0+S_1\right)\right)\nonumber\\
  G_{12}&=&\tau\frac{\gamma}{2}Me^{2i\omega_{s}t_0}\nonumber\\
  G_{21}&=&\tau\frac{\gamma}{2}M^*e^{-2i\omega_{s}t_0}\nonumber\\
  G_{22}&=&-\tau\left(\frac{\gamma}{2}N+iS_1\right).
\end{eqnarray}
If the individual bath modes 
are statistically independent then the real constants $N$ and $S_1$, and
the complex constant $M$
do not depend on the time $t_0$. This would not be the case if some
bath modes with different frequencies were not statistically 
independent.

%
%

\subsubsection{Conditional transition probability in Markov approximation}

\label{T_MA_sec}

In this section, we will combine the results of Sec. \ref{cond_prob}, where
we derived an exact expression for the time evolution of the conditional 
transition probability, 
with the approximations developed in Secs.~\ref{F_calc} and \ref{G_calc}, to
obtain the short time dynamics of the conditional transition probability 
in Markov approximation. 

We shall begin with the evaluation of the correlation matrix of the 
environment $\Gamma$: inserting Eqs.~(\ref{F_app_eq}) and (\ref{G_11_0_eq})
into Eq.~(\ref{Gamm_eq}), we find to first order in $\tau$
\begin{equation}
  \label{corr_first_eq}
  \Gamma=\gamma\tau\left(
  \begin{array}{cc}
    N+1&-Me^{2i\omega_s t_0}\\
    -M^*e^{-2i\omega_s t_0}&N
  \end{array}\right).
\end{equation}
By computing the eigenvalues $\widetilde{\lambda}_i$ and the corresponding 
normalized eigenvectors ${ \mu}_i$ of $\Gamma$, we obtain 
using Eq.~(\ref{J_op_eq}) the jump operators $J_i$. (In order
to simplify the notation, we do not write the $t_0$--dependency of 
$\widetilde{\lambda}_i$ and ${\mu}_i$ explicitly.)

Furthermore, we can calculate the operator $L$, which describes the  
smooth time evolution of the reduced system, by inserting 
Eqs.~(\ref{F_app_eq}) and (\ref{G_11_0_eq}) into Eq.~(\ref{L_eq}). 
The result to first order in $\tau$ is
\begin{equation}
\label{L_MA_eq}
L=I-i\tau\Big(H_{\mbox{\scriptsize Dr}}(t_0)+H_{\rm LS}(t_0)+H_{\rm D}(t_0)\Big),
\end{equation}
where the three operators $H_{\mbox{\scriptsize Dr}}(t_0)$, $H_{\rm LS}(t_0)$, and $H_{\rm D}(t_0)$,
which account for the influence of the environment on the open system
are defined below.
We can distinguish the effects of the two sources of the electromagnetic 
field:

1. The driving field: Using Eq.~(\ref{F_app_eq}) we find
\begin{equation}
  \label{Dr_eq}
  \frac{i}{\tau}\sum_i F_i A_i= i{\cal E}_{t_0}A^\dagger -i{\cal E}_{t_0}^*A
  =: H_{\mbox{\scriptsize Dr}}(t_0).
\end{equation}
This is precisely the operator, which has to be added to the Hamiltonian of
a quantum system in the presence of a classical electromagnetic field 
(according to the principle of minimal substitution) or 
a coherent driving field \cite{Mollow:coh}. However, this result is also true 
for more general driving fields, since the only assumptions we have to 
make about the nature of the driving field concern its correlation times
$\tau_{\cal E}$ and $\tau_{\mbox{\scriptsize Dr}}$ (cf. Secs.~\ref{F_calc} and \ref{G_calc}).

2. The bath: Using Eq.~(\ref{G_11_0_eq}) we find
\begin{equation}
  \label{bath_eq}
  \frac{i}{\tau}\sum_{ij} G_{ij} A_i^\dagger A_j=:H_{\rm LS}(t_0)+H_{\rm D}(t_0), 
\end{equation}
where 
\begin{equation}
  \label{LS_eq}
  H_{\rm LS}=-(S_0+S_1)A^\dagger A+S_1AA^\dagger
\end{equation}
is a Hermitian operator which leads to small energy shifts -- the Lamb shift
$S_0$ and the Stark shift $S_1$. The non-Hermitian operator 
\begin{eqnarray}
  \label{D_eq}
  H_{\rm D}(t_0)&=&-i\frac{\gamma}{2}\bigg((N+1)A^\dagger A+N
  AA^\dagger\nonumber\\
        &-&M^*AAe^{2i\omega_{s}t_0}
        -Me^{-2i\omega_{s}t_0} A^\dagger A^\dagger\bigg)
\end{eqnarray}
describes the damping due to the coupling to the bath. If we define 
$\lambda_i=\widetilde{\lambda}_i/\gamma\tau$, then the damping
operator can also be written as
\begin{equation}
  \label{H_D_1_eq}
  H_{\rm D}(t_0)=-i\frac{\gamma}{2}\sum_i\lambda_i J_i^\dagger J_i.
\end{equation}
Using Eq.~(\ref{H_D_1_eq}) and the fact that $H_{\rm D}(t_0)$ is the only 
non-Hermitian operator which contributes to the operator $L$, 
it is clear that the weight $w=||L\psi_0||^2$ is given by 
\begin{equation}
  \label{w_eq}
  w=1-\gamma\tau\sum_i\lambda_i \langle{\psi_0}|J_i^\dagger 
   J_i|{\psi_0}\rangle=1-\gamma\tau\sum_i \lambda_i w_i.
\end{equation}
The latter equality ensures that $w+\gamma\tau\sum_i\lambda_i w_i=1$, which 
is equivalent to the normalization of the conditional transition probability.

Summing up the above results and transforming back to the Schr\"odinger 
picture, we finally obtain the conditional transition
probability in the Markov approximation
\begin{eqnarray}
  \label{trans_eq}
  T[\psi,t_0+\tau|\psi_0,t_0]&=&\left(1-\gamma\tau\sum_i\lambda_i 
    ||J_i\psi_0||^2\right)\nonumber\\
    &\hspace*{-8em}\times&\hspace*{-4em}
    \delta\left[\psi-\left(I-i\tau G\left(t_0\right)\right)\psi_0 
    \right]\nonumber\\
    &\hspace*{-8em}+&\hspace*{-4em}\gamma\tau\sum_i\lambda_i||J_i\psi_0||^2
  \delta\left[\psi-\frac{J_i\psi_0}{||J_i\psi_0||}\right].
\end{eqnarray}
Here, we introduced the nonlinear time evolution operator \cite{BP:QS2,BP:QS4}
\begin{equation}
  \label{G_eff_eq}
  G(t_0)\psi_0=\left(H_{\rm eff}(t_0)+i\frac{\gamma}{2}\sum_i\lambda_i 
  ||J_i\psi_0||^2\right)\psi_0,
\end{equation}
and the effective system Hamiltonian 
\begin{equation}
\label{H_eff_eq}
H_{\rm eff}(t_0)=H_1(t_0)+H_{\mbox{\scriptsize Dr}}(t_0)+H_{\rm LS}+H_{\rm D}(t_0).
\end{equation}
Note, that the deterministic propagation with the nonlinear operator $G(t_0)$ 
conserves the norm of the propagated state and thus confines the dynamics
to the unit sphere of the Hilbert space of the system.

The following comment may be helpful. If we combine the two approximations 
we made (second 
order perturbation theory and the Markov approximation) with the weak
coupling assumption
and assume that the probability distribution $P[\Psi,t_0]$ of the total
system factorizes at time $t_0$, i.\,e., $P[\cdot,t_0]=P_1[\cdot,t_0]
\otimes P_2[\cdot,t_0]$, then the probability distribution 
$P[\Psi,t_0+\tau]$ of the total system at time $t_0+\tau$ can be approximated
 by
\begin{equation}
  \label{P_t_0+tau_eq}
  P[\cdot,t_0+\tau]=P_1[\cdot,t_0+\tau]\otimes P_2[\cdot,t_0+\tau],
\end{equation}
where $P_1[\cdot,t_0+\tau]$ is given by Eqs.~(\ref{P_prop_1_eq}) and 
(\ref{trans_eq}) and 
$P_2[\cdot,t+\tau]$ by Eq.~(\ref{bath_mo}), i.\,e., $P[\cdot,t_0+\tau]$ 
approximately factorizes. Thus we can use the same scheme
for the calculation of the conditional transition probability
$T[\widetilde\psi,t_0+\tau+\tau'|{\psi},t_0+\tau]$ and by repeating this
argument, we find that $P[\cdot,t]$ factorizes for all $t$. Hence, the
dynamics of the reduced system is Markovian.

%
%

\subsection{Equation of motion for the reduced probability distribution}
\label{eq_mo}

One possible starting point for the derivation of the differential equation 
of motion for the reduced probability distribution
is the infinitesimal generator $\cal G$ of the process \cite{Feller}, which 
is defined by its action on an arbitrary functional $R$, which may depend 
explicitly on time:
\begin{equation}
  \label{generator_eq}
  {\cal G} R[\psi_0,t_0] = \lim_{\tau\rightarrow 0}\frac{{\cal P}_\tau R[\psi_0,t_0]-
    R[\psi_0,t_0]}{\tau},
\end{equation}
where the propagator ${\cal P}_\tau$ is defined by
\begin{equation}
  \label{propagator_eq}
  {\cal P}_\tau R[\psi_0,t_0]= 
  \int D\psi D\psi^* R[\psi,t_0+\tau]T[\psi,t_0+\tau|\psi_0,t_0],
\end{equation}
i.\,e., ${\cal P}_\tau R[\psi_0,t_0]$ is the expectation value of the functional $R$ at 
time $t_0+\tau$ under the condition that the process starts in the state
$\psi_0$ at time $t_0$. Inserting Eqs.~(\ref{trans_eq}) and~(\ref{propagator_eq})
in Eq.~(\ref{generator_eq}) and performing the limit $\tau\rightarrow 0$ 
on the coarse grained time axis (i.\,e., $\tau\gg\tau_{\mbox{\scriptsize B}}$) we obtain
\begin{eqnarray}
  \label{gen_1_eq}
  {\cal G} R[\psi_0,t_0]&=&i\int dx\bigg((G(t_0)\psi_0)^*(x)
  \frac{\delta R[\psi_0,t_0]}{\delta \psi_0^*(x)}\nonumber\\
  &\hspace*{-6em}-&\hspace*{-3em}\frac{\delta R[\psi_0,t_0]}
  {\delta \psi_0(x)}(G(t_0)\psi_0)(x)\bigg)\nonumber\\
  &\hspace*{-6em}+&\hspace*{-3em}\int D\psi^* D{\psi}(R[\psi,t_0]-R[\psi_0,t_0])
  W[{\psi}|\psi_0,t_0],
\end{eqnarray}
where we defined the transition functional
\begin{equation}
  \label{trans_func_eq}
  W[{\psi}|\psi_0,t_0]=\gamma\sum_i \lambda_i 
  \langle{\psi_0}|J_i^\dagger J_i |{\psi_0}\rangle\delta\left[\psi-
  \frac{J_i\psi_0}{||J_i{\psi_0}||}\right].
\end{equation}
Note, that the $t_0$--dependency of $W[{\psi}|\psi_0,t_0]$ is given 
through $\lambda_i$ and $J_i$.
Eq.~(\ref{gen_1_eq}) represents the generator of a piecewise deterministic 
Markov process \cite{Davis}, whose sample paths consist of deterministically 
propagated pieces interrupted by stochastic jumps. The deterministic pieces are
the solution of the nonlinear Schr\"odinger equation \cite{BP:QS2,BP:QS4}
\begin{equation}
  \label{det_eq}
  i\frac{d}{dt}\psi(t)=G(t)\psi(t),
\end{equation}
and the waiting time distribution function for the stochastic jumps is given by
\cite{BP:QS2,BP:QS4}
\begin{equation}
  \label{wait_eq}
  F[\psi_0,t_0,t]=1-\exp\left(-\gamma\sum_i\lambda_i
   \int_{t_0}^t ds ||J_i\psi(s)||^2\right).  
\end{equation}
This is the probability that a jump occurs in the time interval 
$[t_0,t)$ under the condition that the system is in the 
state $\psi_0$ at time $t_0$. If a jump occurs at time $T$ then the
state of the system at time $T$ is
\begin{equation}
\label{Jump_eq}
  |{\psi(T)}\rangle=\frac{J_i|{\psi(T_-)}\rangle}{||J_i|{\psi(T_-)}\rangle||}
\end{equation}
with probability $w_i/(w_1+w_2)$, where $T_-$ is 
$\lim_{\epsilon\rightarrow 0}(T-\epsilon)$, with $\epsilon>0$.

In order to obtain the differential equation of motion for the reduced
probability distribution, we set $R[\psi',t]=\delta[\psi-\psi']$ and 
obtain
\begin{equation}
  \label{trans_2_eq}
  {\cal P}_\tau R[\psi',t]=T[\psi,t+\tau|\psi',t],
\end{equation}
and thus 
\begin{eqnarray}
  \label{p_1_1_eq}
  \frac{\partial}{\partial t}P_1[\psi,t]&=&\lim_{\tau \rightarrow 0}
  \frac{1}{\tau}\Big( P_1[\psi,t+\tau]-P_1[\psi,t]\Big)\nonumber\\
  &\hspace*{-8em}=&\hspace*{-4em}\lim_{\tau \rightarrow 0}\int D\psi'D{\psi'}^*\frac{T[\psi,t+\tau|\psi',t]
  -\delta[\psi-\psi']}{\tau}P_1[\psi',t]\nonumber\\
  &\hspace*{-8em}=&\hspace*{-4em}\int D\psi'D{\psi'}^*\Big({\cal G} \delta[\psi-\psi']
  \Big)P_1[\psi',t].
\end{eqnarray}
If we insert Eq.~(\ref{gen_1_eq}) in Eq.~(\ref{p_1_1_eq}) and evaluate the 
integral we obtain the differential equation of motion of the reduced 
probability distribution
\end{multicols}
\vspace{-0.5cm}
\noindent\rule{0.5\textwidth}{0.4pt}\rule{0.4pt}{\baselineskip}
\widetext
\begin{eqnarray}
  \label{Liou_Mast_eq}
    \frac{\partial}{\partial t}P_1[\psi,t]&=&i\int dx\left\{\frac{\delta}
      {\delta \psi(x)}\Big(G(t)\psi\Big)(x)-\frac{\delta} {\delta \psi^*(x)}
      \Big(G(t)\psi\Big)^*(x)\right\}P_1[\psi,t]\nonumber\\
    &\hspace*{-6em}+&\hspace*{-3em}\int D{\psi'} D{\psi'}^*
    \Big\{W[\psi|{\psi',t}]P_1[{\psi'},t]-W[{\psi'}|\psi,t]P_1[\psi,t]\Big\}
\end{eqnarray}
which is a Liouville--Master Equation \cite{BP:QS2,BP:QS4}. 
Eq.~(\ref{Liou_Mast_eq}) implies that
for the conditional transition probability the integro-differential equation
\begin{eqnarray}
  \label{Liou_Mast_1_eq}
    \frac{\partial}{\partial t}T[\psi,t|\psi_0,t_0]&=&i\int 
      dx\left\{\frac{\delta}
      {\delta \psi(x)}\Big(G(t)\psi\Big)(x)-\frac{\delta} {\delta \psi^*(x)}
      \Big(G(t)\psi\Big)^*(x)\right\}T[\psi,t|\psi_0,t_0]\nonumber\\
    &\hspace*{-10em}+&\hspace*{-5em}\int D{\psi'} D{\psi'}^*
     \Big\{W[\psi|{\psi'},t]T[\psi',t|\psi_0,t_0]-
    W[{\psi'}|\psi,t]T[\psi,t|\psi_0,t_0] \Big\}
\end{eqnarray}
\begin{multicols}{2}
\noindent holds for any $t>t_0$. This equation can be solved either
analytically for some simple systems using the Hilbert--space path integral 
representation of the process \cite{BP:QS8,BP:QS5} or numerically by
either simulating the process \cite{BP:QS2,BP:QS4} or using a
recursive computation scheme \cite{Davis}.

%
%

\section{Reduced Heisenberg picture operators}
\label{heis}

Studying the dynamics of open systems, one is not only interested in the 
time evolution of expectation values of some system operator $X$ 
but, moreover, one wants to investigate the time evolution
of an arbitrary matrix element of this operator and obtain 
quantities such as $\langle{\phi_0,t_0}|X(\tau)|{\psi_0,t_0}\rangle$. 
Note that for an 
open system this is only a short hand notation. What we really mean by this is
\begin{eqnarray}
  \label{Heis_def_eq}
&&  \langle{\phi_0,t_0}|X(\tau)|{\psi_0,t_0}\rangle\nonumber\\
&&\hspace*{3em}\equiv\sum_\alpha p_\alpha
  \langle{\Phi_\alpha,t_0}|e^{iH\tau}\left(X\otimes I\right)e^{-iH\tau}|{\Psi_\alpha,t_0\rangle},
\end{eqnarray}
where $H$ is the Hamiltonian of the total system (see Eq.~(\ref{H_tot_eq}))
and the states $\Phi_\alpha$ and $\Psi_\alpha$ are defined as 
$\Phi_\alpha=\phi_0\otimes\varphi_\alpha$ and  
$\Psi_\alpha=\psi_0\otimes\varphi_\alpha$, i.\,e., we assume, that the state
of the environment is given by the probability distribution 
Eq.~(\ref{env_init_eq}). We will refer to this quantity (Eq.~(\ref{Heis_def_eq}))
as the matrix element of the reduced Heisenberg picture operator $X(\tau)$ 
and use
it in Sec.~\ref{dir_corr} for the
calculation of time-ordered multitime correlation functions.  

In the case of a closed system, $\langle{\phi_0,t_0}|X(\tau)|{\psi_0,t_0}\rangle$ 
is easily calculated by solving 
the corresponding Schr\"odinger equation with the two initial conditions 
$\phi(t_0)=\phi_0$ and $\psi(t_0)=\psi_0$ separately to obtain 
$\phi(t_0+\tau)$ and $\psi(t_0+\tau)$ and then evaluating the scalar product
$\langle{\phi(t_0+\tau)}|X|{\psi(t_0+\tau)}\rangle$. 
A naive generalization of this scheme to an open system -- which will lead
to wrong results -- would be the following: instead of solving the 
Schr\"odinger equation,
simulate the stochastic process defined in Sec.~\ref{eq_mo} with the two 
initial conditions $\phi_0$ and $\psi_0$ to obtain $\phi(t_0+\tau)$ and
$\psi(t_0+\tau)$. Then evaluate the scalar product $\langle{\phi(t_0+\tau)}|
X|{\psi(t_0+\tau)}\rangle$ and average over a sufficiently large ensemble of
realizations. 
However, what we really obtain using this procedure is the quantity 
\begin{eqnarray}
\label{bad_exp_eq}
  f(\phi_0,\psi_0,t_0, \tau) &=&  \int D\phi D\phi^*
  \int D\psi D\psi^*\langle{\phi}|X|{\psi}\rangle\nonumber\\
  &\hspace*{-6em}\times&\hspace*{-3em}
  T[\phi,t_0+\tau|\phi_0,t_0]T[\psi,t_0+\tau|\psi_0,t_0],
\end{eqnarray}
which is, in general, not equal to $\langle{\phi_0,t_0}|X(\tau)|{\psi_0,t_0}\rangle$.
This can be most easily seen by considering the special case 
$\phi_0=\psi_0$: According to the definition of the conditional transition 
probability (cf. Sec.~\ref{cond_prob}), $T[\psi,t_0+\tau|\psi_0,t_0]$ 
has to fulfill the necessary condition 
\begin{eqnarray}
\label{exp_cons_cond_eq}
  \langle{\psi_0,t_0}|X(\tau)|{\psi_0,t_0}\rangle
&=&\int D\psi D\psi^*
  \langle{\psi}|X|{\psi}\rangle \nonumber\\
&\times&  T[\psi,t_0+\tau|\psi_0,t_0].
\end{eqnarray}
Obviously, the quantity $f(\psi_0,\psi_0,t_0,\tau)$ defined in 
Eq.~(\ref{bad_exp_eq})  is, in general, different 
from the right-hand side of Eq.~(\ref{exp_cons_cond_eq}), and thus this 
approach will lead to wrong results.  

However, as we will show below, it is possible to construct
a stochastic process which describes the time evolution of $\phi_0$ and 
$\psi_0$ {\it simultaneously} and whose sample paths can be used to estimate
the desired quantity $\langle{\phi_0,t_0}|X(\tau)|{\psi_0,t_0}\rangle$.
In order to do so, we describe the dynamics of the open system in the 
doubled Hilbert space
$\widetilde{\cal H}_1={\cal H}_1\oplus{\cal H}_1$, such that the ``state'' of 
the open system is given by a normalized pair of state vectors 
$\theta=c^{-1/2}(\phi,\psi)^T$, where $c=||\phi||^2+||\psi||^2$ and $T$ denotes
the transposed vector.
Accordingly, we  define the extension of a system operator $X$ to the doubled 
Hilbert space $\widetilde{\cal H}_1$ as
\begin{equation}
  \label{X_ext_eq}
  \widetilde{X}=\left(\begin{array}{cc}
    I&0\\
    0&X\end{array}\right).
\end{equation}
Using the raising operator
\begin{equation}
  \label{rais_eq}
  S^+=\left(\begin{array}{cc}
    0&I\\
    0&0\end{array}\right),
\end{equation}
whose expectation value $c\langle\theta|S^+|\theta\rangle$ in the state 
$\theta$ is the scalar product $\langle\phi|\psi\rangle$,
we can write the matrix elements of $X$ as
\begin{equation}
  \label{mat_el_exp_eq}
  \langle{\varphi}|X|{\psi}\rangle=c\langle{\theta}|{\widetilde{X}}^\dagger
   S^+\widetilde{X}|{\theta}\rangle.
\end{equation}
This equation is the crucial point of the following construction:
it allows to write matrix elements of reduced system operators
as expectation values in the doubled Hilbert space $\widetilde{\cal H}_1$. 
The same idea will be
used in Sec. \ref{dir_corr} for the development of an algorithm that
enables the determination of arbitrary correlation functions.
Note, that the scalar products on the left-hand side and on the 
right-hand side of the above equations are defined in different Hilbert 
spaces, namely in ${\cal H}_1$ and $\widetilde{\cal H}_1$, respectively.
(The ${\widetilde{X}}^\dagger$ on the right-hand side of 
Eq.~(\ref{mat_el_exp_eq}) is introduced to highlight the connection of the 
above equation with Eq.~(\ref{gen_corr_eq}) (see below)). We also introduce a 
reduced 
probability distribution $\widetilde{P}_1
[\theta,t]$ on the set of normalized states of $\widetilde{\cal H}_1$, 
which is the probability density of finding the system in the state $\theta$,
and a conditional transition probability 
$\widetilde{T}[\theta,t_0+\tau|\theta_0,t_0]$ 
which is the
probability density that the system is in the state $\theta$ at time
$t_0+\tau$ {\it under the condition} that the system is in the state 
$\theta_0$ at time $t_0$.

The goal of this 
section is to derive in analogy to the scheme we presented in Sec. \ref{deri} 
the equation of motion for the conditional transition probability 
$\widetilde T$, which has to fulfill the necessary condition 
\begin{equation}
\label{cons_cond2_eq}
  \langle{\phi_0,t_0}|X(\tau)|{\psi_0,t_0}\rangle=c_0\langle\!\langle 
  {\widetilde{X}}^\dagger
  S^+\widetilde{X}\rangle\!\rangle_{\widetilde{T}[\cdot,t_0+\tau|
  \theta_0,t_0]},
\end{equation}
where $c_0=||\phi_0||^2+||\psi_0||^2$ and $\theta_0=
{c_0}^{-1/2}(\phi_0,\psi_0)^T$. The above condition states that the
matrix element $ \langle{\phi_0,t_0}|X(\tau)|{\psi_0,t_0}\rangle$ of a reduced
Heisenberg picture operator is proportional to 
the expectation value of the operator ${\widetilde{X}}^\dagger
  S^+\widetilde{X}$ with respect to the probability distribution 
$\widetilde T$.
In order to achieve this we have to make the following assumptions, which are 
similar to those made in Sec. \ref{deri}: i) According to the weak coupling 
assumption, we will assume that the   time evolution   of the environment is 
given by its free evolution (cf. Eq. (\ref{bath_mo})). ii) The probability 
distribution of the total system at time $t_0$, $\widetilde{P}[\Theta,t_0]$ 
(which is defined on the Hilbert space $\widetilde{\cal H}=
\widetilde{\cal H}_1\otimes {\cal H}_2$), factorizes and   is given by 
\begin{equation}
  \label{P2_in_eq}
  \widetilde{P}[\Theta,t_0]=\sum_\alpha p_\alpha\int_0^{2\pi}\frac{d\chi}{2\pi}
  \delta[\Theta-e^{i\chi}\theta_0\otimes\varphi_\alpha].
\end{equation}
iii) Second order perturbation theory and finally iv) the Markov approximation.
If we combine these approximations we find after some calculations which 
are similar to
those presented in Sec. \ref{deri}, that the necessary condition 
Eq.~(\ref{cons_cond2_eq}) to first order in $\tau$ (and to second order
in the coupling $g_k$) reads:
\begin{eqnarray}
  \label{cons_cond2_1_eq}
  \langle{\phi_0,t_0}|X(\tau)|{\psi_0,t_0}\rangle&=&
  \langle{\phi_0}| L^\dagger X L|{\psi_0}\rangle\nonumber\\
  &+& \gamma\tau\sum_{i}\lambda_i\langle{\phi_0}| J_i^\dagger X J_i
  |{\psi_0}\rangle\nonumber\\
  &\stackrel{!}{=}&c_0\langle\!\langle {\widetilde{X}}^\dagger S^+\widetilde{X}
  \rangle\!\rangle_{\widetilde{T}[\cdot,t_0+\tau|\theta_0,t_0]},
\end{eqnarray}
where $L$, and $\lambda_i$, and $J_i$ are defined as in Sec.~\ref{T_MA_sec}. 
Note that this equation is completely analogous to Eq.~(\ref{exp_prop_dia}).
This leads us directly to a conditional transition probability to first order 
in $\tau$
\begin{eqnarray}
\label{Heis_T_eq}
{\widetilde{T}[\theta,t_0+\tau| \theta_0,t_0]}&=& \widetilde{w}\delta[
  \theta-\widetilde{w}^{-1/2}\widetilde{L}\theta_0]\nonumber\\
  &+&\gamma\tau\sum_{i}\lambda_i
  \widetilde{w}_i\delta[\theta-\widetilde{w}_i^{-1/2}\widetilde{J}_i\theta_0],
\end{eqnarray}
where we have defined
\begin{equation}
  \label{L_J_ext_eq}
   \widetilde{H}_{\rm eff}(t)=\left(\begin{array}{cc}
    H_{\rm eff}(t)&0\\
    0&H_{\rm eff}(t)\end{array}\right),\quad
  \widetilde{J}_i=\left(\begin{array}{cc}
    J_i&0\\
    0&J_i\end{array}\right).
\end{equation}
The effective Hamiltonian $H_{\rm eff}(t)$ is defined in Eq.~(\ref{H_eff_eq}), and 
$\widetilde{L}=I-i\tau\widetilde{H}_{\rm eff}$. The appropriate normalization 
constants are $\widetilde{w}=||\widetilde{L}\theta_0||^2$ and 
$\widetilde{w}_i=||\widetilde{J}_i\theta_0||^2$.
In analogy to Eq.~(\ref{w_eq}) we find
\begin{equation}
  \label{w_tilede_eq}
  \widetilde{w}=1-\gamma\tau\sum_i\lambda_i
  ||\widetilde{J}_i\theta_0||^2=1-\gamma\tau\sum_i \lambda_i\widetilde{w}_i,
\end{equation}
which ensures the normalization of the conditional 
transition probability $\widetilde{T}$. 

The equation of motion for the conditional transition probability 
$\widetilde{T}[\theta,t|\theta_0,t_0]$, where $t>t_0$, 
 can be found using the same scheme as described in 
Sec.~\ref{eq_mo}: Calculation of the generator ${\cal G}$ and its action on
the functional $R[\theta',t]=\delta[\theta-\theta']$
leads to the Liouville--Master equation 
\end{multicols}
\widetext
\begin{eqnarray}
  \label{Heis_Liou_Mast_eq}
    \frac{\partial}{\partial t}
    \widetilde{T}[\theta,t|\theta_0,t_0]
    &=&i\int dx\bigg\{\frac{\delta}{\delta \theta(x)}
    \cdot\Big(\widetilde{G}(t)\theta\Big)(x)
    -\frac{\delta} {\delta \theta^*(x)}\cdot\Big(\widetilde{G}(t)\theta\Big)^*(x)
    \bigg\}\widetilde{T}[\theta,t|\theta_0,t_0]\nonumber\\
    &\hspace*{-8em}+&\hspace*{-4em}\int D{\theta'} D{\theta'}^*\Big\{
    W[\theta|\theta',t]
    \widetilde{T}[\theta',t|\theta_0,t_0]
   -W[\theta'|\theta,t]
   \widetilde{T}[\theta,t|\theta_0,t_0]\Big\},
\end{eqnarray}
\begin{multicols}{2}
\noindent where we defined the functional Wirtinger derivative with respect to states 
belonging to $\widetilde{\cal H}_1$ as
\begin{equation}
  \label{func_der_eq}
  \frac{\delta}{\delta \theta(x)}=\left(\begin{array}{c}
      \displaystyle\frac{\delta}{\delta \phi(x)}\\[2ex]\displaystyle
      \frac{\delta}{\delta \psi(x)}
  \end{array}\right).
\end{equation}
The generator for the continuous time evolution is defined
in analogy to Eq.~(\ref{G_eff_eq}) as
\begin{equation}
\label{Heis_G_eq}
  \widetilde{G}(t)\theta=\left(\widetilde{H}_{\rm eff}(t)
  +i\frac{\gamma}{2}\sum_i\lambda_i
  ||\widetilde{J}_i\theta||^2\right)\theta,
\end{equation}
and the transition functional is given by (cf.~Eq.~(\ref{trans_func_eq}))
\begin{equation}
  \label{Heis_trans_func_eq}
  W[\theta|\theta_0,t]=\gamma\sum_i \lambda_i 
  \widetilde{w}_i\delta\left[\theta-\widetilde{w}_i^{-1/2}
  \widetilde{J}_i\theta_0\right].
\end{equation}
Again, the stochastic process which is defined through 
Eq.~(\ref{Heis_Liou_Mast_eq}) is a piecewise deterministic Markov process, where
the deterministic pieces are the solution of the nonlinear Schr\"odinger 
equation
\begin{equation}
\label{Heis_det_eq}
i\frac{d}{dt}\theta=\widetilde{G}(t)\theta
\end{equation}
and the waiting time distribution for the stochastic jumps under the condition
that the system is in the state $\theta_0$ at time
$t_0$ is given by
\begin{equation}
  \label{Heis_wait_eq}
  F[\theta_0,t_0,t]=1-
  \exp\left(-\gamma\sum_i\lambda_i\int_{t_0}^tds||\widetilde{J}_i\theta(s)||^2
   \right).
\end{equation}
The Liouville-Master equation (\ref{Heis_Liou_Mast_eq}) also determines the
equation of motion for the reduced Heisenberg picture operator $X(t)$: 
Expanding $X(t)$ in terms of its matrix elements and
inserting the Liouville-Master equation (\ref{Heis_Liou_Mast_eq})
we obtain after some calculations
\begin{eqnarray}
  \label{op_eq_mo_eq}
  \lefteqn{\frac{d}{dt}X(t) = i[H_1+H_{\rm LS}+H_{\rm Dr},X](t)
  + \frac{1}{2}\sum_i\gamma_i }\nonumber\\
  &\times&\Big\{
  2\left(J_i^\dagger X J_i\right)(t)-\left(J_i^\dagger J_i X\right)(t)-\left(X J_i^\dagger J_i\right)(t)
  \Big\}.
\end{eqnarray}
Thus the equation of motion of the reduced Heisenberg picture operator is the 
adjoint equation of the quantum master equation \cite{Alicki,Sondermann} and
we find
\begin{equation}
  \label{reg_bas_eq}
  \mbox{Tr}_{\mbox{\scriptsize sys}}\Big\{XV(t,t_0)\{\rho_0\}\Big\}=
  \mbox{Tr}_{\mbox{\scriptsize sys}}\Big\{X(t)\{\rho_0\}\Big\}
\end{equation}
for any initial density matrix $\rho_0$, where $V(t,t_0)$ is the time 
evolution super operator of the quantum master equation. This relation is the 
basis of the quantum regression theorem \cite{Walls,GardinerQN}.

We close this section with a few remarks: For the  initial condition 
$\theta_0=(\psi_0,\psi_0)^T/\sqrt{2}$ the stochastic process 
defined in this way reduces to
the stochastic process defined in Sec.~\ref{eq_mo}; formally, this can
be expressed through the identity
\begin{equation}
\label{Heis_red_eq}
\widetilde{T}[\theta,t|\theta_0,t_0]=\delta\left[\theta-\frac{1}{\sqrt{2}}
(\psi,\psi)^T\right]
T[\psi,t|\psi_0,t_0],
\end{equation}
where $T[\psi,t|\psi_0,t_0]$ is the 
solution of the Liouville--Master equation (\ref{Liou_Mast_eq}). Thus,
the dynamics of the system described in the doubled Hilbert space 
$\widetilde{\cal H}_1$ reduces to our earlier description of the dynamics
in the ordinary Hilbert space ${\cal H}_1$. This is an important property
of the conditional transition probability $\widetilde{T}$ which establishes
the link between the two approaches. 

For the numerical simulation of the stochastic process, it is useful 
to omit the normalization requirement and use the linear operator 
$\widetilde{H}_{\rm eff}$
instead of $\widetilde{G}$ as the generator for the deterministic motion. 
Thus, for the numerical simulation we use an unnormalized state 
vector $\hat{\theta}$, and the deterministic pieces are the
solution of the {\it linear} Schr\"odinger-type equation
\begin{equation}
\label{Heis_lin_eq}
 i\frac{d}{dt}\hat\theta=\widetilde{\cal H}_{\rm eff}\hat\theta
\end{equation}
and the waiting time distribution for the stochastic jumps for unnormalized 
states is given by
\begin{equation}
  \label{Heis_wait_unnorm_eq}
  F[\theta_0,t_0,t]=1-||\hat{\theta}(t)||^2,
\end{equation}
under the condition that the state $\theta_0$ is normalized and $\hat\theta(t)$
is the solution of the above Schr\"odinger-type equation with the initial 
condition $\hat\theta(t_0)=\theta_0$. This equation
is easily proven by differentiating Eqs.~(\ref{Heis_wait_eq}) and 
(\ref{Heis_wait_unnorm_eq}) with respect to $t$ and comparing the results.
The above procedure has the major advantage, that the Schr\"odinger equation 
(\ref{Heis_lin_eq}) for $\hat{\theta}=(\hat\phi,\hat\psi)$ is linear and 
that the two components are decoupled. This leads to a large reduction of 
the time necessary for the calculation of the deterministic 
time evolution. In addition, we do not have to calculate the waiting
time distribution separately.

In Ref. \cite{Gisin:Heis} Gisin formulated the Heisenberg picture
of the quantum state diffusion model of open systems. In analogy to our 
approach, he used a coupled pair of stochastic differential equations
for the calculation of Heisenberg picture operators. His approach has
the advantage that the scalar product of two state vectors
$\phi$ and $\psi$ is constant in time, i.\,e., the matrix elements of 
the identity operator are calculated correctly in each realization of
the stochastic process, whereas in our approach the identity operator 
is treated like any other system operator, i.\,e., for the calculation of
the matrix elements one has to average over many realizations. 
On the other hand, Gisin's approach is limited to the calculation of
matrix elements with respect to nonorthogonal state vectors, i.\,e.,
the stochastic differential equations are not defined if $\langle \phi
|\psi\rangle =0$. Furthermore, the quasi linear stochastic equations 
proposed for the numerical simulation, are not decoupled -- in contrast
to our approach -- which is a disadvantage from a numerical point
of view. 

%
%

\section{Calculation of multitime correlation functions}
\label{mult_corr}

\subsection{Symmetric time-ordered multitime correlation functions}

\label{symm_corr}
The theory we presented in Sec. \ref{deri} allows the calculation of
one time expectation values of an observable such as 
$\langle{\psi_0,t_0}|X(\tau)|{\psi_0,t_0}\rangle$. 
This theory can easily be extended to the calculation of symmetric, 
time-ordered multitime correlation functions such as 
\begin{equation}
\label{symm_corr2_eq}
g(\psi_0,t_0,t_1,t_{2})=\langle{\psi_0,t_0}|X^\dagger(t_1)Y(t_{2})X(t_1)|{\psi_0,t_0}\rangle,
\end{equation}
where $X$ and $Y$ are arbitrary operators and  $t_0\le t_1\le t_{2}$.
As in Sec.~\ref{heis} we use the short hand notation 
\begin{eqnarray}
  \label{symm_corr2_1_eq}
  \langle{\psi_0,t_0}|X^\dagger(t_1)Y(t_{2})X(t_1)|{\psi_0,t_0}\rangle
  \nonumber\\
  &\hspace*{-22em}\equiv&\hspace*{-10em}
  \sum_\alpha p_\alpha 
  \langle{\Psi_{\alpha},t_0}|X^\dagger(t_1)Y(t_{2})X(t_1)
  |{\Psi_{\alpha},t_0}\rangle,
\end{eqnarray}
where ${\Psi_{\alpha}}=\psi_0\otimes\varphi_\alpha$ is a pure product
state of the total system. Thus, at time $t_0$ the system is in the pure state
$\psi_0$ and the state of the environment is given through the probability 
distribution $P_2$ (cf. Eq.~(\ref{env_init_eq})). An exact expression for
$g(\psi_0,t_0,t_1,t_{2})$ in terms of probability 
distributions defined on the Hilbert space of the {\it total} system 
can be written as
\begin{eqnarray}
\label{symm_corr2_2_eq}
  g(\psi_0,t_0,t_1,t_{2})&=&\int D\Psi_1 D\Psi_1^*\int D\Psi_2 D\Psi_2^*
  \langle{\Psi_{2}}|Y|{\Psi_{2}}\rangle\nonumber\\
  &\hspace*{-10em}\times&\hspace*{-5em} \langle{\Psi_1}|X^\dagger X|{\Psi_{1}}\rangle
  \hat{T}\bigg[\Psi_2,t_2\bigg|\frac{X\Psi_1}{||X\Psi_1||},t_1\bigg]
  P[\Psi_1,t_1].
\end{eqnarray}
The time-dependent probability distribution $\hat{P}$ and the conditional 
transition probability $\hat{T}$ for the total system can be calculated using 
the unitary time evolution, 
\begin{eqnarray}
\label{symm_corr2_2a_eq}
&&  P[\Psi_1,t_1]=\sum_{\alpha}p_\alpha
  \delta[\Psi_1-\exp(-iH(t_0-t_1)) \psi_0\otimes
  \varphi_\alpha]\nonumber\\
&&  \hat{T}[\Psi_2,t_2|\Psi_1,t_1]=
  \delta[\Psi_2-\exp(-iH(t_2-t_1))\Psi_1],
\end{eqnarray}
where $H$ is the Hamilton of the total system (cf. Eq~(\ref{H_tot_eq})).
If we apply the scheme we presented in Sec.~\ref{deri} (i.\,e., the weak
coupling assumption, second order perturbation theory, the reduction formula, 
and the Markov approximation) 
to express $P$ and $\hat{T}$ in terms of the conditional transition probability
$T$ of the reduced system we find that $g(\psi_0,t_0,t_1,t_{2})$ is 
approximately given by
\begin{eqnarray}
\label{symm_corr2_3_eq}
g(\psi_0,t_0,t_1,t_{2})&=&\int D\psi_1 D\psi_1^*\int D\psi_2 D\psi_2^*
\langle{\psi_{2}}|Y|{\psi_{2}}\rangle\nonumber\\
&\hspace*{-12em}\times&\hspace*{-6em}
  ||X\psi_1||^2 \;T\bigg[\psi_2,t_2\bigg|\frac{X\psi_1}{||X\psi_1||},t_1\bigg]
T[\psi_1,t_1|\psi_0,t_0],
\end{eqnarray}
where the $\psi_i$ are states of the system under consideration and the
conditional transition probability $T$ satisfies the 
Liouville--Master Equation Eq.~(\ref{Liou_Mast_1_eq}). For the stochastic
process which is used to simulate Eq.~(\ref{symm_corr2_3_eq}) this means the
following: 
the process starts in the state $\psi_0$ at time $t_0$ and is propagated
to the state $\psi_1$ at time $t_1$
using the stochastic time evolution defined in Sec.~\ref{cond_prob}.
Then the process jumps to the normalized state $X\psi_1/||X\psi_1||$ and is 
propagated to the state $\psi_2$ at time $t_2$. 

For the most general
symmetric time-ordered correlation function we obtain in a similar way
\end{multicols}
\vspace{-0.5cm}
\noindent\rule{0.5\textwidth}{0.4pt}\rule{0.4pt}{\baselineskip}
\widetext
\begin{eqnarray}
\label{symm_corr_eq}
g(\psi_0,t_0,...,t_n,t_{n+1})&=&\langle{\psi_0,t_0}|X_1^\dagger(t_1)\cdots X_n^\dagger(t_n)Y(t_{n+1})
X_n(t_n)\cdots X_1(t_1)|{\psi_0,t_0}\rangle\nonumber\\
   &\hspace*{-14em}=&\hspace*{-7em}\int D\psi_1 D\psi_1^*\cdots
   \int D\psi_{n+1} D\psi_{n+1}^*\langle{\psi_{n+1},t_{n+1}}|Y
   |{\psi_{n+1},t_{n+1}}\rangle w_1\cdots w_n\times\nonumber\\
   &\hspace*{-14em}\times&\hspace*{-7em} T[\psi_{n+1},t_{n+1}|{w_n}^{-1/2}X_n 
   \psi_{n},t_{n}]\cdots 
   T[\psi_2,t_2|{w_1}^{-1/2}X_1\psi_1,t_1]T[\psi_1,t_1|\psi_0,t_0],
\end{eqnarray}
\begin{multicols}{2}
\noindent where $X_i$ and $Y$ are arbitrary system operators, and  
$w_i=||X_i\psi_i||^2$, and $t_0\le t_1\le\cdots\le t_{n+1}$. This class of 
correlation functions contains
for example the correlation $\langle{\sigma^+(t)\sigma^+(t+\tau)
\sigma^-(t+\tau)\sigma^-(t)}\rangle$ which has been measured in the famous 
Hanbury-Brown and Twiss experiment (see, e.~g. \cite{Hanbury:HBT}) or, more
recently, for a single ion in a trap by Diedrich and Walther 
\cite{Walther:HBT}. 

The stochastic simulation algorithm 
which we use to compute symmetric time-ordered multitime correlation 
functions can thus be summarized as follows: 
1.  Start in the state $\psi_0$ at time $t_0$ and propagate up to
  the state $\psi_1$ at time $t_1$ using the stochastic time evolution 
   (cf. Sec.~\ref{eq_mo}).
2. Jump to the state $X_1\psi_1/||X_1\psi_1||$. 
3. Propagate the state $X_1\psi_1/||X_1\psi_1||$ at time $t_1$
  up to the time $t_2$, etc. 
4. Compute the expectation value $\langle\psi_{n+1},t_{n+1}| Y|
   \psi_{n+1},t_{n+1}\rangle$.
By repeating this procedure sufficiently often, we can estimate the 
unknown correlation function by averaging over the sample paths.

In the preceding discussion we limited ourselves to a pure
initial state $\psi_0$ of the system. However, this is not a real restriction:
if the initial state of the open system is described by a probability 
distribution $P_1[\psi,t_0]$, we find
\begin{eqnarray}
  \label{mix_init_corr_eq}
&&  \langle\!\langle X_1^\dagger(t_1)\cdots X_n^\dagger(t_n)Y(t_{n+1})
  X_n(t_n)\cdots X_1(t_1)\rangle\!\rangle_{P_1[\cdot,t_0]}\nonumber\\
&&  \hspace*{2em}=\int D\psi D\psi^*
  P_1[\psi,t_0] g(\psi,t_0,...,t_n,t_{n+1})
\end{eqnarray}
Thus, we can simulate this type of correlation functions by drawing 
a random initial state $\psi$ with probability $P_1[\psi,t_0]$ and using
the above simulation algorithm.

%
%

\subsection{General time-ordered multitime correlation functions}
Especially in the context of quantum optics, measurements are not restricted 
to obtain information about symmetric time-ordered correlation functions.
For example, the fluorescence spectrum of a two level atom is the
Fourier transform of the stationary expectation value 
$\langle\!\langle{\sigma^+(\tau)\sigma^-}\rangle\!\rangle_{s}$ which
is a special case of the more general time-ordered multitime correlation 
function
\begin{eqnarray}
\label{corr_eq}
&&g(\psi_0,t_0,t_1,...,t_n,s_1,...,s_m)\nonumber\\
&&\hspace*{1.5em} =\langle{\psi_0,t_0}|X_1^\dagger(t_1)\cdots 
X_n^\dagger(t_n)Y_m(s_m)\cdots Y_1(s_1)|{\psi_0,t_0}\rangle,\nonumber\\
\end{eqnarray}
where $t_0\le\cdots\le t_n$, and $t_0\le s_1\le\cdots\le s_m$, and
$X_i$ and $Y_i$ are arbitrary system operators. 
In fact, according to Gardiner \cite{GardinerQN} and Gardiner and Collett 
\cite{Gard:QSDE}, this kind of multitime correlation function is the only 
one that arises in the quantum theory of measurement. 

For the calculation of this time-ordered correlation function we will present
two distinct approaches: we can express the general time-ordered correlation
function through a linear combination of symmetric time-ordered 
correlation functions (Sec. \ref{red_corr}) or we can define a stochastic 
process in the doubled Hilbert space ${\cal H}_1\oplus{\cal H}_1$ 
(cf.~Sec.~\ref{heis}) which allows
a direct calculation of the sought quantity (Sec.~\ref{dir_corr}).

%
%

\subsubsection{Reduction to symmetric time-ordered correlation functions
 in ${\cal H}_1$}
\label{red_corr}

In order to express the general time-ordered correlation function defined in 
Eq.~(\ref{corr_eq}) as a linear combination of symmetric time-ordered 
correlation functions as defined in Eq.~(\ref{symm_corr_eq}), we can use the 
operator identity 
\begin{eqnarray}
\label{pol_ident_eq}
&&X^\dagger MY=\frac{1}{N}\sum_{k=1}^Ne^{-2\pi ki/N}\nonumber\\
&& \hspace*{.5em}{\left(X+e^{2\pi ki/N}Y\right)}^\dagger M\left(X+e^{2\pi ki/N}Y\right),\quad N\ge 3
\end{eqnarray}
(which yields the polarization identity \cite{GardinerQN} 
for $N=4$). 
This identity is 
easily proven by computing the product and recognizing that the prefactor
of every term in the sum is a geometric series. 
For example the correlation function $\langle\!\langle X^\dagger(t+\tau)
Y(t)\rangle\!\rangle$ could be rewritten as
\begin{eqnarray}
  \label{Moel_eq}
&&  \langle\!\langle X^\dagger(t+\tau)Y(t)\rangle\!\rangle=
  \frac{1}{4}\sum_{k=1}^4e^{-2\pi ki/4}\nonumber\\
&&\hspace*{.5em}\Big\langle\!\Big\langle
  \left(I+e^{2\pi ki/4}Y(t)\right)^\dagger X^\dagger(t+\tau)
  \left(I+e^{2\pi ki/4}Y(t)\right)\Big\rangle\!\Big\rangle.\nonumber\\
\end{eqnarray}
This is a symmetric time-ordered correlation function. Relation 
(\ref{Moel_eq}) was 
used by Dalibard et~al. \cite{MolmerPRL68} for the computation of the
spectrum of fluorescence of a two level atom.
However, for the 
calculation of a general time-ordered correlation function, one would
need to calculate at least $3^{n+m-r-s-1}$ symmetric time-ordered correlation 
functions, where $r$ is the number of indices $i$ for which $t_i$ is equal to 
some $s_j$ and $s$ is the number of indices $i$ for which $t_i$ and ${X_i}$
are equal to some $s_j$ and $Y_j$, respectively. Especially for
higher order correlation functions, this procedure is very time consuming.

%
%

\subsubsection{Calculation by the stochastic process in the doubled Hilbert
space ${\widetilde{\cal H}_1}$}
\label{dir_corr}

In Sec.~\ref{symm_corr}, we demonstrated how the stochastic process
which we use to simulate one--time expectations of system operators 
(cf. Sec.~\ref{eq_mo})
could be applied to the simulation of symmetric time-ordered multitime 
correlation functions. Similarly, the stochastic process which we use
to calculate matrix elements of a reduced Heisenberg picture operator
$\langle{\phi_0,t_0}|X(\tau)|{\psi_0,t_0}\rangle$ (cf. Sec.~\ref{heis}) 
provides a direct method for the calculation of general time-ordered 
correlation functions of the type given by Eq.~(\ref{corr_eq}). The reason
for this fact is---as we will show below---that these correlation 
functions can be expressed by symmetric correlation functions in the doubled 
Hilbert space 
$\widetilde{\cal H}_1$ in the following manner:
Order the set of times $\{t_1,\cdots t_n,s_1,\cdots s_m\}$ and rename
 them $r_i$ such
that $r_1<\cdots<r_{n+m-k}$ where $k$ is the number of coinciding time points.
 Then define a set of Schr\"odinger operators $F_l$ 
on $\widetilde{\cal H}_1$ as
\begin{equation}
  \label{F_l_eq}
 F_l=\left\{\begin{array}{cl}
  \left(\begin{array}{cc}
    X_i&0\\
    0&I\end{array}\right),
       & \mbox{if $r_l=t_i\neq s_j$ for some $i$ and all $j$,}\\[3ex]
  \left(\begin{array}{cc}
    I&0\\
    0&Y_j\end{array}\right),
       & \mbox{if $r_l= s_j\neq t_i$ for some $j$ and all $i$,}\\[3ex]
  \left(\begin{array}{cc}
    X_i&0\\
    0&Y_j\end{array}\right),
       & \mbox{if $r_l=t_i= s_j$ for some $i$ and $j$}.
\end{array}\right.
\end{equation}
Using this definitions and the definition of the raising operator 
$S^+$ (cf. Eq.~(\ref{rais_eq})) we find  from Eq.~(\ref{corr_eq})
\begin{eqnarray}
  \label{gen_corr_eq}
&&  g(\psi_0,t_0,t_1,...,t_n,s_1,...,s_m)\nonumber\\
&&\hspace*{1.5em}=c_0
 \langle{\theta_0}|F_1^\dagger(r_1)\cdots F_q^\dagger(r_{q})S^+
  F_{q}(r_{q})\cdots F_1(r_1)|{\theta_0}\rangle,\nonumber\\
\end{eqnarray}
where $q=n+m-k$, $c_0=2||\psi_0||^2$, and $\theta_0=c_0^{-1/2}
(\psi_0,\psi_0)$. Eq.~(\ref{gen_corr_eq}) is a symmetric time-ordered multitime 
correlation function that can be calculated applying the simulation algorithm 
in a doubled Hilbert space described in 
Sec.~\ref{symm_corr}. 
Note that for the special choice $n=0$, $m=1$, $t_1=t_0$, 
$c_0=||\phi||^2+||\psi||^2$, and $\theta_0=c_0^{-1/2}(\phi,\psi)$ the 
right-hand side of Eq.~(\ref{gen_corr_eq}) is just the matrix element 
$\langle \phi|Y|\psi\rangle$ of some system operator $Y$ 
(cf. Eq.~(\ref{mat_el_exp_eq})).

As an explicit example of the general formalism we will consider the 
two time correlation function 
$\langle{\psi_0,t_0}|X^\dagger(t_1)Y(t_2)|{\psi_0,t_0}\rangle$, with
$t_0\le t_1\le t_2$: for this we obtain the following simulation algorithm:
1. Start in the normalized state $\psi_0$ at time $t_0$ and propagate
 to the state $\psi_1$ at time $t_1$ using the stochastic time evolution
 in ${\cal H}_1$ (cf. Sec.~\ref{eq_mo}).
2. Jump to the normalized pair of state vectors 
$c_1^{-1/2}(X\psi_1,\psi_1)$, where $c_1=||X\psi_1||^2+||\psi_1||^2$.
3. Propagate to the state $(\phi_2,\psi_2)$ at time $t_2$ using the 
stochastic time evolution defined in the doubled Hilbert space 
(cf. Sec.~\ref{heis}).
4. Evaluate the scalar product $\langle{\phi_{2}}|Y|{\psi_{2}}\rangle$ and
weight it with the factor $c_1$.
Repeat this procedure sufficiently often and average over the different 
realizations. Using Eq.~(\ref{reg_bas_eq}) it is straightforward to show 
that applying this scheme we determine the quantity
\begin{equation}
  \label{reg_corr_eq}
  \mbox{Tr}_{\rm sys}\bigg\{YV(t_2,t_1)\Big\{V(t_1,t_0)
  \{|\psi_0\rangle\langle\psi_0|\}X^\dagger\Big\}\bigg\},
\end{equation}
where $V(t',t)$ is the propagator of the quantum master equation. This is 
in complete agreement with the standard definition of the two time correlation 
function $\langle{\psi_0,t_0}|X^\dagger(t_1)Y(t_2)|{\psi_0,t_0}\rangle$ 
\cite{GardinerQN} and, hence, in complete agreement with the quantum regression
theorem.

 The above algorithm is easily generalized to the calculation of higher order 
correlation functions. Note that by using this approach we have to 
calculate only one 
symmetric time-ordered multitime correlation function in a doubled
Hilbert space, instead of $3^{n+m-r-s-1}$ correlation functions that have
to be calculated using the method described in Sec.~\ref{red_corr}. 
We will compare the numerical
performance of both methods in Sec.~\ref{vac} where we use the calculation 
of the fluorescence spectrum of a driven two level atom as an example.

%
%

\section{Examples}
\label{examples}
\subsection{Coherently driven two level atom}
 
\label{vac}

Consider a two level atom with the Hamiltonian $H_1=\omega_s\sigma^+\sigma^-$,
where $\sigma^\pm$ are the pseudo spin 
operators of the atom, driven by a coherent field. The state of the 
environment is given by
\begin{equation}
  \label{vac_env_state_eq}
  |{\varphi}\rangle=\prod_k D(\beta_ke^{-i\omega t})|{0}\rangle
\end{equation}
with probability 1, where $D$ is the unitary displacement operator 
and $\beta_k$ is the amplitude of the coherent excitation of the field mode 
$k$.  We will assume that the bandwidth of the coherent excitation 
is small, i.\,e., Eq.~(\ref{tau_tau_E_cond}) holds.
Using Eq.~(\ref{E_t_eq}) we find for the mean of the electromagnetic field
\begin{equation}
  \label{E_vac_eq}
  {\cal E}_{t}=-i\langle\!\langle B_2\rangle\!\rangle_{P_{2}[\cdot,t]}=-i\sum_{k\in{\cal K}_{\mbox{\scriptsize Dr}}}
  g_k\beta_k e^{-i\omega_k t},
\end{equation}
and the driving term in the effective Hamiltonian is thus given by
\begin{equation}
  \label{drive_vac_eq}
  H_{\mbox{\scriptsize Dr}}=\sum_{k\in{\cal K}_{\mbox{\scriptsize Dr}}}g_k\left(\beta_k e^{-i\omega_k t}\sigma^++
  \beta_k^*\sigma^-e^{i\omega_k t}\right).
\end{equation}
Since the bath is in the vacuum state, we find using Eqs.~(\ref{n_eq}) and
(\ref{m_eq}) from the Appendix
\begin{equation}
\label{n_m_vac_eq}
\overline{n}(\omega_{s})=\overline{m}(\omega_s)=0.
\end{equation}
Computing the eigenvalues of the correlation matrix of the environment 
$\Gamma$ (cf.~Eq.~(\ref{corr_first_eq})) leads to 
$\lambda_1=1$, and $\lambda_2=0$, and thus we find the jump 
operators $J_1=\sigma^-$, and $J_2=\sigma^+$. Neglecting the Lamb shift
we obtain for the effective Hamiltonian 
\begin{eqnarray}
  \label{H_eff_vac_eq}
  H_{\rm eff}&=&\left(\omega_s-i\frac{\gamma}{2}\right)\sigma^+\sigma^-\nonumber\\
&+&\sum_{k\in{\cal K}_{\mbox{\scriptsize Dr}}}g_k
  \left(\beta_k e^{-i\omega_k t}\sigma^++
  \beta_k^*\sigma^-e^{i\omega_k t}\right).
\end{eqnarray}
This is the same effective Hamiltonian, that is used in other 
stochastic wave function approaches
for the coherent time evolution of a system coupled to the vacuum, see, e. g. 
Refs. \cite{MolmerPRL68,ZollerPRA46,BP:QS8}.

\begin{figure}[p]
  \begin{center}
    \leavevmode \epsfxsize\linewidth\epsffile{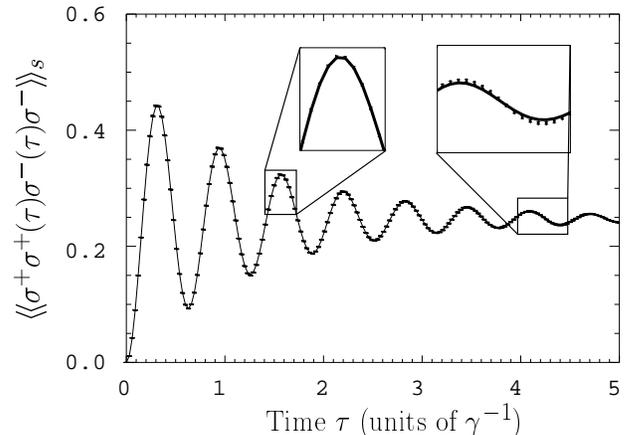}
    \caption{\narrowtext Correlation function $\langle\!\langle \sigma^+\sigma^+(\tau)
      \sigma^-(\tau)\sigma^-\rangle\!\rangle_{s}$ for a
      coherently driven two level atom on resonance with Rabi frequency 
      $\Omega=10\gamma$. The
      numerical result for $10^5$ realizations of the simulation algorithm
      described in Sec.~\protect{\ref{symm_corr}} (errorbars) 
      and analytical solution (thick line).}
    \label{ana_HBT}
  \end{center}
\end{figure}

We can now use the simulation algorithm described in Sec.~\ref{symm_corr}
to calculate the correlation function 
$\langle\!\langle \sigma^+(t)\sigma^+(t+\tau)
\sigma^-(t+\tau)\sigma^-(t)\rangle\!\rangle_{s}$ which is the probability of 
emitting a 
photon at time $t$ and a second photon at time $t+\tau$ in the steady state. 
This correlation
function  could be measured in a Hanbury-Brown and Twiss experiment. We do 
this by 
drawing a random initial state from an uniform distribution on ${\cal H}_1$ 
and then propagating this state for a time $t=30\gamma^{-1}$ in order to 
reach the steady state regime. Then we apply the above described simulation 
algorithm to calculate the sought quantity.
In Fig.~\ref{ana_HBT} we compare the numerical results with the analytical
solution obtained by the quantum regression theorem (solid line) for a 
resonant coherent excitation with 
${\cal E}_t=5\gamma e^{-i\omega_s t}$ (which yields the Rabi frequency
$\Omega=10\gamma$) and $10^5$ realizations.
The error bars indicate the estimated error calculated at each 
point using the estimated standard deviation of the sample of realizations. 
Obviously, our calculations are in good agreement with the analytical results.

\begin{figure}[p]
  \begin{center}
    \leavevmode  \epsfxsize\linewidth\epsffile{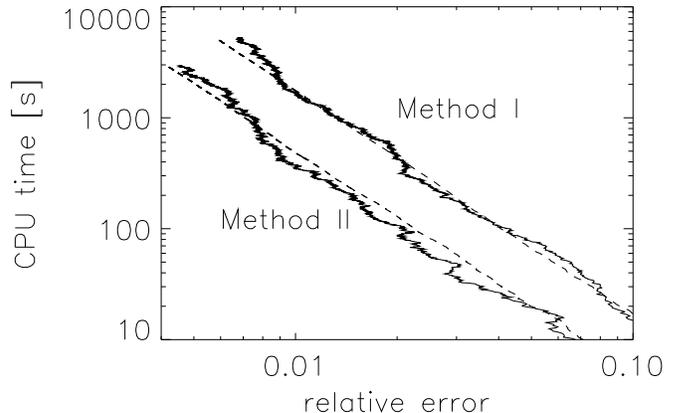}  
    \caption{\narrowtext Calculation of the first order correlation function 
$\langle\!\langle \sigma^+(\tau)\sigma^-\rangle\!\rangle_{s}$ for a 
coherently driven two level atom on resonance. This figure shows the CPU time 
in seconds  vs.~the relative error for the simulation algorithms proposed in 
\protect\cite{MolmerPRL68} (Method I) 
and for our algorithm (cf. Sec.~\protect{\ref{dir_corr}}, Method II). 
The solid lines represent the mean square deviation of the numerical solution 
from the exact solution and the dashed lines show the estimated standard 
deviation of the numerical solution.}
\label{num_per}
\end{center}
\end{figure}

We can also compute the spectrum of the fluorescence radiation which is 
the Fourier transform of the correlation function 
$\langle\!\langle\sigma^+(t+\tau)\sigma^-(t)\rangle\!\rangle$  
using the simulation algorithms presented in 
Sec.~\ref{red_corr} (denoted by Method I in Fig.~\ref{num_per}) and 
Sec.~\ref{dir_corr} (denoted by Method II in  Fig.~\ref{num_per}) respectively.
For the same parameters as above, both algorithms
converge to the analytical solution, first given by Mollow \cite{Mollow:spec}.
However, the numerical performance of both algorithms is
quite different: Fig.~\ref{num_per} shows the necessary CPU time on an 
RS6000 workstation in
order to achieve a desired numerical precision in a log--log
plot. The solid lines represent the mean square deviation of the
numerical solution from the exact solution and the dashed lines show
the mean estimated standard deviation of the numerical solution.
Obviously, the latter quantity provides for both algorithms a very
good measure of the accuracy of the numerical simulation.
A closer analysis reveals, that Method II is for a given accuracy 
by a factor of $3$ faster than Method I. Actually, this result is better 
than we expected, since the time Method I uses for the computation of a single 
realization is only by a factor of 1.7 greater than the time Method II
needs. Thus, the accuracy of a single realization is higher for Method II than
for Method I. We emphasize that, in order to obtain a realistic picture of the
scaling of the CPU times, we used a numerical integration of the corresponding
 differential equations (Eqs.~(\ref{det_eq}) and (\ref{Heis_det_eq})), 
instead of the well-known analytical solution. 

%
%

\subsection{Two level atom in a squeezed vacuum}
\label{squeeze}

As a second example, we want to investigate a two level atom with the 
Hamiltonian ${\cal H}_1=\omega_s\sigma^+\sigma^-$, driven by a displaced
squeezed vacuum. Thus, the state of the environment is given by 
\begin{equation}
  \label{sq_env_state_eq}
  |{\varphi}\rangle=\prod_k  D(\beta_ke^{-i\omega_k t})S(r_ke^{-2i\phi_k})
  |{0}\rangle
\end{equation}
with probability 1, where $D$ is the unitary displacement operator 
and $S$ is the unitary squeezing operator. Using the identities \cite{Walls}
\begin{eqnarray}
  \label{sq_ident_eq}
  S^\dagger(re^{-2i\phi})b_kS(re^{-2i\phi})&=&b_k\cosh r 
  - b_k^\dagger e^{-2i\phi}\sinh r\nonumber\\
  S^\dagger(re^{-2i\phi})b_k^\dagger S(re^{-2i\phi})&=&b_k^\dagger 
  \cosh r - {b_k} e^{2i\phi}\sinh r
\end{eqnarray}
together with Eq.~(\ref{E_t_eq}) we find for the mean of the electromagnetic 
field
\begin{equation}
  \label{E_sq_eq}
  {\cal E}_{t}=-i\langle\!\langle B_2\rangle\!\rangle_{P_{2}[\cdot,t]}=-i\sum_{k\in{\cal K}_{\mbox{\scriptsize Dr}}}
  g_k\beta_k e^{-i\omega_k t},
\end{equation}
and thus for the driving term in the effective Hamiltonian 
\begin{equation}
  \label{drive_sq_eq}
  H_{\mbox{\scriptsize Dr}}=\sum_{k\in{\cal K}_{\mbox{\scriptsize Dr}}}g_k\left(\beta_k e^{-i\omega_k t}\sigma^++
  \beta_k^*\sigma^-e^{i\omega_k t}\right).
\end{equation}
It is important to note that the driving term is completely unaffected by the
squeezing and again the coherent excitation acts like a classical field. If we 
assume that the bandwidth of the coherent excitation is
small and that the squeezing is homogeneous over a wide bandwidth near
the resonance frequency $\omega_s$ then Eq.~(\ref{Mark_cond_eq}) holds and
we can calculate the parameters $N$ and $M$ appearing in the bath correlation
matrix $\Gamma$ using Eqs.~(\ref{n_eq}) and (\ref{m_eq}). We find
\begin{equation}
  \label{N_M_sq_eq}
  N=\varepsilon\sinh^2r_s,\quad M=\varepsilon e^{-2i\phi_s}\sinh r_s\cosh r_s,
\end{equation}
where $r_s$ and $\phi_s$ are the parameters of the squeezing at the resonance
frequency $\omega_s$ and the efficiency parameter $\varepsilon$ is defined as
$\varepsilon=|\sigma|/4\pi$ where $\sigma$ is the
solid angle in which the squeezing occurs. If the squeezing is homogeneous
over the complete solid angle $4\pi$, then $\varepsilon=1$ and we find a 
perfect squeezed vacuum with $|M|^2=N(N+1)$ (minimum uncertainty 
state); if only a finite solid angle is squeezed, then we find 
$|M|^2=N(N+\varepsilon)$, which corresponds to an imperfect squeezing. 
Thus a perfect squeezing over a finite
solid angle leads to the same dynamics as an imperfect squeezing
within the complete solid angle $4\pi$. 

In order to obtain the rates $\gamma\lambda_i$ and the corresponding jump 
operators $J_i$ which are necessary for a stochastic description of the 
system we have to calculate the eigenvalues and eigenvectors of the bath 
correlation matrix $\Gamma$. In terms of $N$ and $\epsilon$ we find 
\begin{eqnarray}
  \label{lam_j_sq_eq}
  \lambda_{1,2}&=&N+\textstyle{\frac{1}{2}}\pm \sqrt{N(N+\varepsilon)+\textstyle{\frac{1}{4}}}\nonumber\\
  J_1&=&\cos\theta e^{i\phi_s}A-\sin\theta e^{-i\phi_s}
  A^\dagger\nonumber\\
  J_2&=&\sin\theta e^{i\phi_s}A+\cos\theta e^{-i\phi_s}
  A^\dagger,
\end{eqnarray}
where $\tan2\theta=2\sqrt{N(N+\epsilon)}$. 

\begin{figure}[p]
  \begin{center}
    \leavevmode  \epsfxsize\linewidth\epsffile{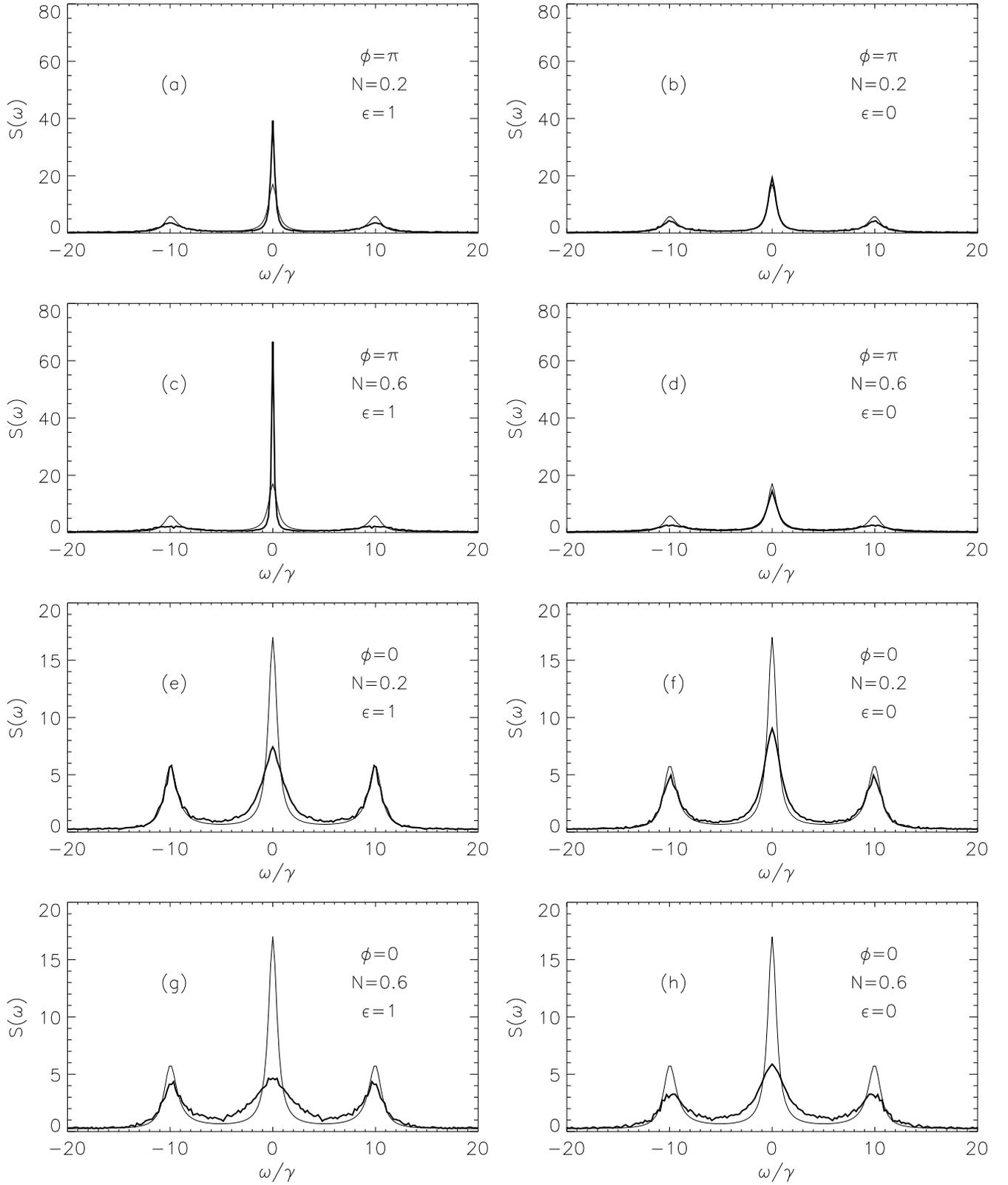}  
    \caption{Resonance fluorescence spectra for a coherently driven two level 
     atom with Rabi frequency $\Omega=10\gamma$ coupled to a squeezed vacuum 
     for different squeezing parameters (mean photon number $N$, relative phase
     of squeezing and driving field $\phi=2(\phi_s-\phi_L)$, and fraction
     $\varepsilon$ of solid angle which is squeezed): numerical results
     (solid line), and the vacuum spectrum (thin line). The numerical solution
     was obtained by simulating a stochastic process in a doubled Hilbert 
     space (cf. Sec.~\protect{\ref{dir_corr}}).}
   \label{sq_specs}
\end{center}
\end{figure}

The spectrum of 
fluorescence can be obtained by calculating the correlation function
$\langle\!\langle\sigma^+(\tau)\sigma^-\rangle\!\rangle_{s}$ in the 
steady state, and then taking the Fourier transform of this quantity. 
In Fig.~\ref{sq_specs} we present 
numerically calculated spectra of resonance fluorescence for 
a coherent driving field with a Rabi frequency of $\Omega=10\gamma$ 
(${\cal E}_t=5\gamma e^{-i(\omega_L t-\phi_L)}$) and various
squeezing parameters (solid lines) and compare it with the 
vacuum spectrum obtained by Mollow \cite{Mollow:spec} (thin line).
In Ref.~\cite{Carmichael87} it was shown that in the
strong field limit the spectrum of resonance
fluorescence is sensitive to the relative phase of the driving field
and the phase of the squeezing, i.\,e., it depends on $\phi=2(\phi_s-\phi_L)$:
For a perfect squeezing and $\phi=0$ the linewidth of the central peak
is reduced to a subnatural level, whereas the sideband peaks are broadened.
This is shown in Fig.~\ref{sq_specs} (a) and \ref{sq_specs} (c). In the 
limit $\varepsilon 
\rightarrow 0$ but $N=\mbox{const}.$ (i.\,e., very strong squeezing of only a few
modes) the shape of the central peak is approximately identical with the 
shape of the 
central peak in the Mollow triplet; the sideband peaks are still broadened
(cf. Fig.~\ref{sq_specs} (b) and \ref{sq_specs} (d)).
On the other hand, for $\phi=\pi$ the central peak and the sideband peaks 
are broadened for both $\varepsilon =1$ 
(cf. Fig.~\ref{sq_specs} (e) and \ref{sq_specs} (g)) and $\varepsilon =0$ 
(cf. Fig.~\ref{sq_specs} (f) and \ref{sq_specs} (h)). 

For the numerical calculation of the correlation function  
$\langle\!\langle\sigma^+(\tau)\sigma^-\rangle\!\rangle_{s}$
we used the stochastic simulation 
algorithm in the doubled Hilbert space described in Sec.~\ref{dir_corr} 
and $10^4$ realizations. Note, that it
is useful from a numerical point of view to subtract the 
limit $\lim_{\tau\rightarrow\infty}\langle\!\langle\sigma^+(\tau)
\sigma^-\rangle\!\rangle_{s}$ from the calculated correlation function
before taking the Fourier transform, since this constant corresponds to
the $\delta-$shaped coherent part of the spectrum, which is neglectable
in the limit of strong driving fields. The agreement of the numerical with the
analytical solution given in Ref.~\cite{Carmichael87} is very good for all 
the examples presented above (the relative error of the correlation function 
is of the order $10^{-2}$). Note that the analytical solution 
was obtained by assuming that all modes of the electromagnetic field
are squeezed. However, the above discussion makes clear that it can also
be used for the more general case considered here by setting 
$|M|^2=N(N+\varepsilon)$, i.\,e., by assuming an imperfect squeezing. 

%
%

\subsection{Thermal mixture of coherent states}
\label{therm}

As a last example we want to investigate an open system coupled to a 
heat bath in a thermal state, which is represented by a mixture of coherent 
states. Thus the probability density 
$P_{\mbox{\scriptsize B}}[\varphi,t_0]$ of the bath is given by 
\begin{equation}
  \label{therm_env_eq}
  P_{\mbox{\scriptsize B}}[\cdot,t_0]=\bigotimes_{k\in{\cal K}_{\mbox{\scriptsize B}}}P_k[\cdot,t_0],
\end{equation}
where 
\begin{eqnarray}
  \label{coh_therm_eq}
  P_k[\varphi,t_0]&=&\frac{1}{\pi}\left(e^{\omega_k/kT}-1\right)\int d^2\alpha
  \int_0^{2\pi}\frac{d\chi}{2\pi}\nonumber\\
   &\hspace*{-6em}\times&\hspace*{-3em}\exp\left(-|\alpha|^2\left(e^{\omega_k/kT}
   -1\right)\right)\delta\left[\varphi-e^{i\chi}|\alpha\rangle\right],
\end{eqnarray}
and $|\alpha\rangle$ is an eigenstate of the destruction operator $b_k$ with
complex eigenvalue $\alpha$. By inserting Eq.~(\ref{coh_therm_eq}) in
Eqs.~(\ref{n_eq}) and (\ref{m_eq}), respectively, we obtain
\begin{equation}
  \label{N_M_eq}
  N=\frac{1}{e^{\omega_k/kT}-1}, \quad M=0,
\end{equation}
as expected, and hence the jump operators $J_i=A_i$ and rates $\gamma
\lambda_1=\gamma(N+1)$ and  $\gamma\lambda_2=\gamma N$. Note that we would 
have obtained the same values for
$N$ and $M$ and hence the same equation of motion for the reduced
probability distribution 
by representing the probability density $P_{\mbox{\scriptsize B}}$ of the heat bath 
in terms of eigenstates of the number operator \cite{BP:QS2,BP:QS4}. 
The reason for this invariance is that all parameters which describe
the influence of the environment on the open system are expectation
values of linear environment operators and hence independent of the
actual representation of the mixture. 

%
%

\section{Summary}
The primary goal of the present paper was to show how 
stochastic wave function methods have to be generalized
in order to allow the systematic evaluation of the matrix elements
of reduced Heisenberg operators and of arbitrary multitime
correlation functions. 
The starting point for the derivation of the equation of motion for the
reduced system's probability distribution is the unitary time evolution
of the total system (i.\,e., the system together with its environment). 
From this, we infer the dynamics of the reduced
system through the following necessary condition: the time evolution of the 
expectation value of an arbitrary system observable $C$ calculated using
the reduced  probability distribution has to be equal to the time evolution
of the expectation value of $C\otimes I$ in the total system (up to
second order perturbation theory). By making the week coupling assumption
and the Markov approximation we can further simplify the expression
for the reduced system's probability distribution and finally obtain the
desired equation of motion. This equation is a Liouville--Master equation
and hence the underlying stochastic process is a piecewise deterministic
Markov process. We emphasize, that the only major assumptions
we have to make about the state of the environment concern the relation of
the time scale of the reduced system's dynamics compared to the time scale 
of the dynamics of the environment.
  
The basic idea underlying the approach to the evaluation of multitime 
correlations presented here is the introduction of a doubled
Hilbert space. The latter allows to write matrix elements of
reduced Heisenberg picture operators as expectation values. Employing this
ansatz a stochastic process in the doubled Hilbert space is easily
constructed which 
directly leads to the determination of arbitrary
correlation functions. The process in the doubled Hilbert space
is again a piecewise deterministic process described by a
Liouville--Master equation for a probability distribution on the 
doubled Hilbert space.

To complete this article we have discussed three typical examples of quantum
optics: (i) A coherently driven two level system coupled to the vacuum.
This example is mainly used as a basis for a numerical comparison 
of two simulation algorithms for the calculation of multitime correlation 
functions, namely our approach and an algorithm proposed in 
Ref.~\cite{MolmerPRL68}. For this example our approach was faster by a factor 
of 3. (ii)  A coherently driven two level system coupled 
to a squeezed vacuum. In this example we derive the stochastic time evolution
of a system coupled to a squeezed vacuum which extends over a finite solid
angle (i.\,e., not over the complete solid angle $4\pi$). In particular
we show, that a perfect squeezing over a finite solid angle leads to the
same dynamics of the open system as an imperfect squeezing over the complete 
solid angle $4\pi$.
(iii) A thermal mixture of coherent states. This example illustrates, that
the equation of motion for the reduced probability distribution does not
depend on the actual representation of the state of the environment.

%
%

\begin{appendix}
\section{Calculation of the ensemble average of the second order
  propagation operator}
\label{G_ij_app}

Let us calculate first $G_{22}$:
Combining Eq.~(\ref{G_lin_eq}) and~(\ref{bath_correl_eq})
and performing the continuum limit yields 
\begin{equation}
  \label{G_22_eq}
  G_{22}=-\tau\frac{\gamma}{2\pi}\int_{0}^\infty d\omega_k
  \overline{n}(\omega_k)\int_0^\infty ds\; e^{-i(\omega_s-\omega_k)s},
\end{equation}
where we defined the mean number of photons $\overline{n}(\omega_k)$ with 
frequency $\omega_k$ as
\begin{equation}
  \label{n_eq}
  \overline{n}(\omega_k)=\frac{2\pi\omega^2V}{(2\pi c)^3\gamma}\sum_{\lambda_k}
  \int d\Omega\; g_k^2\langle\!\langle b_k^\dagger b_{k}
  \rangle\!\rangle_{P_{k}[\cdot,0]},
\end{equation}
where $k=(\omega_k,\hat{\mbox{\boldmath $k$}},\lambda_k)$ and
\begin{equation}
  \label{k_hat_eq}
  \hat{k}(\Omega)=\left(\begin{array}{c}
  \sin\theta\cos\varphi\\
  \sin\theta\sin\varphi\\
  \cos\theta\end{array}\right),
\end{equation}
and the mean quadratic coupling at frequency $\omega_s$ 
\begin{equation}
  \label{gamm_eq}
  \gamma=\frac{2\pi\omega_s^2V}{(2\pi c)^3}\sum_{\lambda_k}
  \int d\Omega\; g_{(\omega_s,\hat{k},\lambda_k)}^2.
\end{equation}
Note that the above summations and integrations are restricted to modes $k\in
{\cal K}_{\mbox{\scriptsize B}}$. Since $g_k^2\sim V^{-1}$ the expressions for 
$\overline{n}(\omega_k)$ and $\gamma$ are independent of the volume of 
quantization $V$. Using the identity
\begin{eqnarray}
  \label{int_ident_eq}
&&  \frac{1}{2\pi}\int_{-\infty}^\infty d\omega_k f(\omega_k)
  \int_0^\infty d\tau e^{i\omega_k \tau} \nonumber\\
&&\hspace*{2em}=\frac{1}{2}f(0)+\frac{i}{2\pi}
  \mbox{P}\int_{-\infty}^\infty d\omega_k f(\omega_k)/\omega_k,
\end{eqnarray}
where P denotes the Cauchy principal value we find 
\begin{equation}
  \label{G_22_eq_2}
  G_{22}=-\tau(\frac{\gamma}{2}N+iS_1),
\end{equation}
with $N=\overline{n}(\omega_s)$ and 
\begin{equation}
  \label{stark_eq}
  S_1=\frac{\gamma}{2\pi}\mbox{P}\int_{-\infty}^\infty d\omega
  \frac{\overline{n}(\omega_s+\omega)}{\omega}.
\end{equation}
Thus, the constant $N$ we have just defined is the mean number of 
photons in the modes with frequency $\omega_s$, where the mean is taken over
the states of the ensemble {\it and} over the relative strength of the 
coupling. The imaginary part $S_1$ leads to the Stark shift.

For the calculation of $G_{11}$ we can use
the commutator relations $[b_k,b_{k'}^\dagger]=\delta_{kk'}$ and find:
\begin{equation}
  \label{G_11_eq}
  G_{11}=G_{22}^*-\tau\frac{\gamma}{2\pi}\int_{-\infty}^{\infty}d\omega_k
h(\omega_k)\int_0^\infty ds e^{i(\omega_{s}-\omega_k)s},
\end{equation}
where 
\begin{equation}
  h(\omega_k)=\frac{2\pi\omega^2V}{(2\pi c)^3\gamma}\sum_{\lambda_k}
  \int d\Omega\; g_k^2.
\end{equation}
Thus we find 
\begin{equation}
  \label{G_11_1_eq}
  G_{11}=-\tau(\frac{\gamma}{2}(N+1)-i(S_0+S_1))
\end{equation}
where we used the definition of $\gamma$ (Eq.~(\ref{gamm_eq})) and defined
\begin{equation}
  \label{lamb_eq}
  S_0=\frac{\gamma}{2\pi}\mbox{P}\int_{-\infty}^\infty d\omega_k
  \frac{h(\omega_s+\omega_k)}{\omega_k}.
\end{equation}
The new constant $S_0$ which appears here is the Lamb shift.

In a similar way we can also calculate $G_{12}=G_{21}^*$ by defining 
\begin{equation}
  \label{m_eq}
  \overline{m}(\omega)=\frac{2\pi\omega^2V}{(2\pi c)^3\gamma}\sum_{\lambda_k}
  \int d\Omega\; g_k^2\langle\!\langle{b_{k}} b_{k}
  \rangle\!\rangle_{P_{k}[\cdot,0]},
\end{equation}
where the summation and integration extend over all modes $k\in{\cal K}_{\mbox{\scriptsize B}}$, and
\begin{equation}
  M=\overline{m}(\omega_s)+\frac{i}{\pi}\mbox{P}\int_{-\infty}^\infty d\omega
  \frac{\overline{m}(\omega_s+\omega)}{\omega},
\end{equation}
which leads to the result
\begin{eqnarray}
  \label{corr_1_eq}
  G_{12}&=&\tau\frac{\gamma}{2}M e^{2i\omega_{s}t_0}\\
  G_{21}&=&\tau\frac{\gamma}{2}M^*e^{-2i\omega_{s}t_0}.
\end{eqnarray}

\end{appendix}

\bibliographystyle{prsty}  

\end{multicols}

\widetext
\end{document}